# Protein Structure Prediction by Protein Alignments

By

Jianzhu Ma

Submitted to:
**Toyota Technological Institute at Chicago**
6045 S. Kenwood Ave, Chicago, IL, 60637

September 2015

For the degree of Doctor of Philosophy in Computer Science

Thesis Committee:

| | | |
|---|---|---|
| Jinbo Xu (Thesis Supervisor) | Signature: | Date: Oct 14, 2015 |
| Tobin Sosnick | Signature: | Date: Oct 20, 2015 |
| Karen Livescu | Signature: | Date: October 13, 2015 |
| Matthew Walter | Signature: | Date: October 13, 2015 |

# Abstract

Proteins are the basic building blocks of life. They form the basis of hormones, which regulate metabolism, structures such as hair, wool, muscle, and antibodies. In the form of enzymes, they are behind all chemical reactions in the body. They also help our body fight infections, turn food into energy, copy DNA and catalyze chemical reactions. In fact, 60% of the average human body is water and 17% is proteins.

Proteins usually perform their functions by folding to a particular structure. Understanding the folding process could help the researchers to understand the functions of proteins and could also help to develop supplemental proteins for people with deficiencies and gain more insight into diseases associated with troublesome folding proteins. A lot of efforts have been devoted to develop the experimental methods to determine the structure 3D structure of proteins, such as X-ray crystallography and NMR spectroscopy. However, experimental methods are both expensive and time consuming.

In this thesis I try to introduce a new machine learning template protein structure prediction problem to predict the protein structure. The new method improves the performance from two directions: creating accurate protein alignments and predicting accurate protein contacts.

For the first direction, the thesis presents an alignment framework MRFalign which goes beyond state-of-the-art methods and uses Markov Random Fields (MRFs) to model a protein family and align two proteins by aligning two MRFs together. Compared to other methods, that can only model local-range residue correlation, MRFs can model long-range residue interactions (e.g., residue co-evolution) and thus, encodes global information in a protein. MRFalign formulates the problem as an integer programming problem and use an Alternative Direction Method of Multipliers (ADMM) algorithm to quickly find a suboptimal alignment of two MRFs.

For the second direction, the thesis presents a Group Graphical Lasso (GGL) method for contact prediction that integrates joint multi-family Evolutionary Coupling (EC) analysis and supervised learning to improve accuracy on proteins without many sequence homologs. Different from existing single-family EC analysis that uses residue co-evolution information in only the target protein family, our joint EC analysis uses residue co-evolution in both the target family and its related families, which may have divergent sequences but similar folds.

Our GGL method can also integrate supervised learning methods to further improve accuracy.

We evaluate the performance of both methods including each of its components on large public benchmarks. Experiments show that our methods can achieve better accuracy than existing state-of-the-art methods under all the measurements on most of the protein classes.

# Acknowledgements

First, let me thank my adviser Jinbo Xu for guiding me through my graduate studies. He has been encouraging and supportive of my research from beginning to end. I can't remember how many times I went into his office with nearly every crazy idea and he listened to me patiently and helped me formalize them into concrete directions to pursue. He has tremendous insight into what approaches will succeed. He also shaped my view of looking at things and taught me the real spirit of research, which I believe will be beneficiary to my whole life, and that I will cherish forever.

I would like to thank my committee members Tobin Sosnick, Karen Livescu and Matthew Walter for all their support and suggestions. My heartfelt thanks are also dedicated to all the professors from TTI-C and University of Chicago who have taught me, David McAllester, John Lafferty, Nati Srebro, Julia Chuzhoy, Karen Livescu, Greg Shakhnanovich, Yury Makarychev, Aly Khan, Qixing Huang, Shi Li, Hammad Naveed, Stefan Canzar, for their encouraging and sharing their plethora of knowledge on various subjects and courses. I also thank many other professors and colleagues who had left TTI-C, Tamir Hazan, Raquel Urtasun, Yang Shen and Dhruv Batra.

During the 5 years in TTI-C, I have made lots of friends. I thank for their friendship and constantly support: Jian Peng, Sheng Wang, Hao Tang, Karthik Sridharan, Andrew Cotter, Zhiyong Wang, Avleen Bijral, Payman Yadollahpour, Siqi Sun, Qingming Tang, Lifu Tu, Jian Yao, Xing Xu and Somaye Hashemifar. Thank you for your good company and support.

I also thank all the Administrative Staff of TTI-C for helping deal with lots of personal staffs: Chrissy Novak, Anna Ruffolo, Adam Bohlander and Liv Leader. Sorry for giving you so much trouble and thank you for always being so patient to me.

Finally, my biggest thanks go to my loving and supportive family. This thesis is dedicated to my beloved wife, Chao Yue. Thank you for sharing this journey with me. It was because of you that my work becomes meaningful.

# Contents

# Chapter 1 Introduction

In a cell, proteins carry out various types of biological functions by folding into particular 3D structures. Thus, elucidating a protein's structure is the key to understanding its function, which in turn is essential for any further related biological, medical, or pharmaceutical applications. Currently, experimental determination of a protein structure is still expensive, labor intensive and time consuming. The gap between the number of available protein sequences and the number of proteins with experimentally determined structures is still large. Fortunately, computational methods for predicting protein structures can partially solve this problem and provide biologists with valuable information on the proteins they are interested in. Among all these computational methods, different statistical machine learning methods have been proposed over the years and have significantly contributed to advancing the state-of-the-art in protein structure prediction.

There are two major approaches to predict the structure of a protein, 1) Template-based Modeling (also called comparative modeling or protein threading) and 2) Template-Free modeling (also called ab initio modeling or free folding). Template-based modeling methods use the previously determined protein crystal structures similar to the query protein to predict the structure for it. This technique is based on the fact that proteins with similar sequences or evolutionary traces tend to have similar structures. Template-free modeling seeks to predict the structure of a protein from the protein sequence alone by minimizing a particular kind of energy function. It is based on the belief that the native structures of most proteins correspond to the ones with the lowest free energy. When homologous templates can be found, template-based method are more reliable compared to template-free method. In this thesis we will mainly focus on template-based modeling, although some of the techniques for contact prediction may also help the template-free method.

Template-based modeling has three steps, as shown in Figure 1: 1) Align the query protein to each of the protein in the template database; 2) Select one or several templates based on the evolutionary and structural features calculated from the corresponding alignments; 3) Build 3D structures for the target protein considering the constraints provided by the aligned regions while at the same time minimizing a particular energy of the unaligned loop regions and add side-chain atoms. The last two steps can be merged into a single procedure since one could select the templates based on the quality of recovered 3D structure. The bottleneck of template-based modeling following these steps is the accuracy of alignments. Both template selection and 3D structure recovery rely on accurate alignment between target protein and templates. Most popular template selection methods use features from each aligned and unaligned position from the given alignments and used them for their own ranking functions. The quality of the alignments will consequently influence the quality of the features they fetched. In the 3D structure recovery, the modeling software will take the alignments as constraints to restrict the optimization of their own energy function. An incorrect alignment might lead to the optimization to a completely wrong conformation space. Therefore, improving the alignment quality will help both of these tasks in template-based modeling. In this thesis, I will focus on developing new probabilistic graphical models for protein alignment including both new energy functions and new structures.

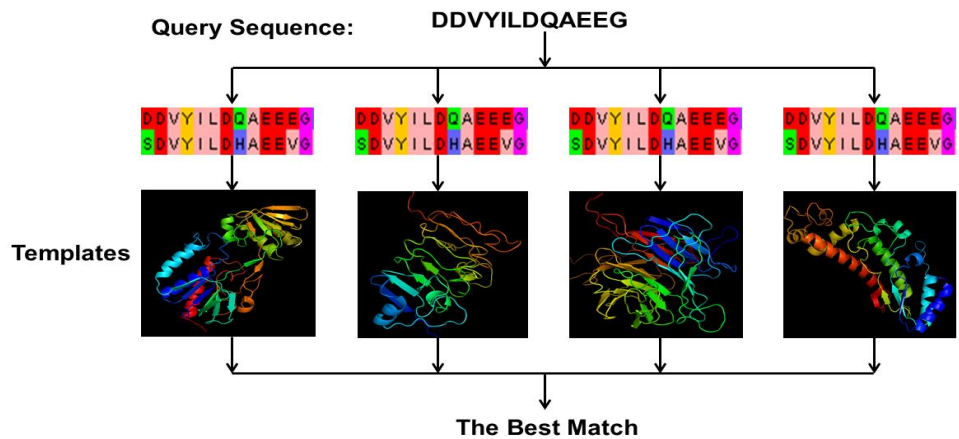

**Figure 1.** Pipeline for template-based modeling for protein structure prediction.

In a machine learning perspective, protein alignment can be treated as a structured prediction problem with the goal to predict the alignment state for each pair of the residues from the two proteins to be aligned. The alignment state for a pair of residue is related to their local features. For example, if residues $i$ and $j$ on two proteins have the same amino acid type, similar mutation frequency and secondary structure type, then they are more likely to be aligned together. The alignment states are not independent with each other. For example, the alignment state of residues $i$ and $j$ depends on the alignment states of their adjacent residues. Similar as other structured learning problem, the key point of solving this problem is to design a computational model that can well capture both of these two dependencies. In order to capture the first dependency, we need to design a new scoring function that is expressive enough to model the complex relationship between features and alignment states. For the second dependency, we need to design a graphical model with new graph structure with new training and inference methods.

The thesis is organized as follows: in Chapter 2, I will introduce the background knowledge of protein structures and various protein alignment approaches. In Chapter 3, I will describe a new graphical model-based protein alignment method, which can be applied to solving both protein homology detection and protein threading. In Chapter 4, I will introduce a novel alignment scoring function that can capture the complex relationship between the protein features and alignment states. In Chapter 5, I will introduce a novel protein alignment potential function. In Chapter 6, I will introduce a new contact prediction approach that can be treated as a new feature used in our alignment framework. In Chapter 7, I will conclude the thesis and discuss the future work.

# Chapter 2  Protein structure and protein alignment

## 2.1 Protein Structure

Protein structure is usually described at four different levels as shown in Figure 2. Each of the lower levels can be treated as the building blocks of the higher level. The hierarchical structure of proteins is very important for prediction. The lower level structure is usually much easier to predict compared to the higher level and we can therefore predict the overall 3D structure from bottom-up.

The first level, called the primary sequence, is a linear sequence of the amino acids in the chain. Different primary structures correspond to different sequences in which the amino acids are covalently linked together. Amino acids are organic compounds composed of amine (-NH2) and carboxylic acid (-COOH) functional groups, along with a side-chain specific to each amino acid. There are 20 types of standard amino acids altogether. During a protein folding process, amino acids are connected by the chemical bonds through a reaction of their respective carboxyl and amino groups. These bonds are called peptide bonds and the amino acids linked by the peptide bonds are called peptides, or residues.

The second level, called the secondary structure, has two common patterns of structural repetition in proteins: the coiling up of segments of the chain named $\alpha$-helix, and the pairing together of strands of the chain named $\beta$-sheet. These two structures are more conserved compared to other regions, which are usually referred to as coil or loop. The coil region is important for maintaining the flexibility and binding affinity when the protein interacts with others. Instead of using this definition, Sander grouped the secondary structure into eight classes (Kabsch and Sander, 1983). This classification is a finer-grained model of the 3-classes one and contains more useful information, such as the difference between 3-helix and 4-helix.

The tertiary structure is the next higher level of organization. It is defined as the set of 3D coordinates for each atoms of the protein. The folding of the polypeptide chain assembles different secondary structure elements in a particular arrangement. As helices and sheets are units of secondary structure, the domain is the unit of tertiary structure that is a conserved part of a given protein sequence and (tertiary) structure that can evolve, function, and exist independently of the rest of the protein chain. In multi-domain proteins, tertiary structure includes the arrangement of domains relative to each other as well as that of the chain within each domain.

The quaternary structure describes how different polypeptide chains are assembled into complexes.

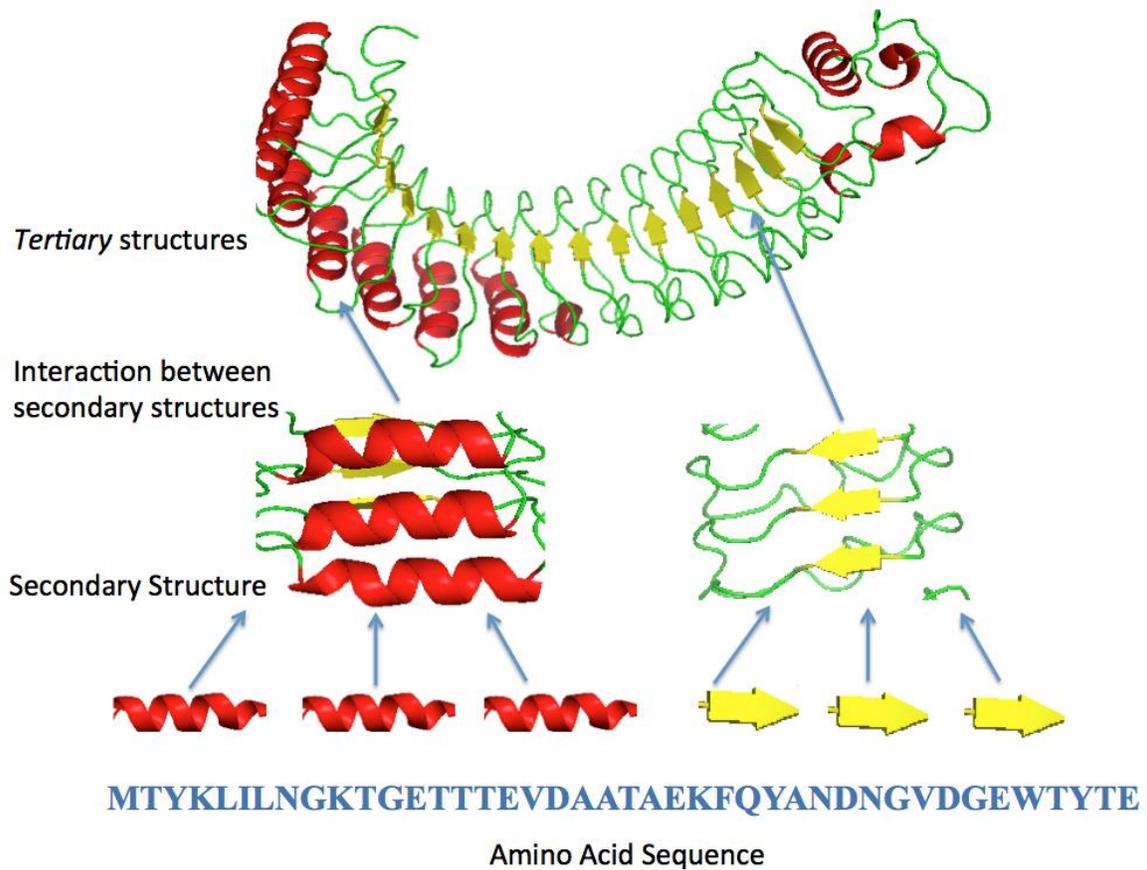

**Figure 2.** Different layers of protein structures.

## 2.2 Protein Alignment Methods

As mentioned in the previous section, the most reliable protein structure prediction method is template-based method. Its bottleneck is the quality of protein alignments. According to the features of proteins used for different methods under study, alignment-based methods can be grouped into three categories: sequence-based alignment methods, profile-based alignment methods and structure-based methods. Generally speaking, sequence-based alignment methods are less sensitive than profile-based alignment methods, which in turn are less sensitive than structure-based alignment methods. However, sequence-based methods are more specific than profile-based alignment methods, which in turn are more specific than structure-based alignment methods.

Sequence-based methods can build relatively accurate alignments for close homologous proteins. A few methods have been developed and their difference mainly lies in alignment algorithms, amino acid mutation score and gap penalty. Some methods such as the Needleman-Wunsch (Needleman and Wunsch, 1970) and Smith-Waterman algorithms (Smith and Waterman, 1981) employ dynamic programming to build alignments, while others such as BLAST (Altschul, et al., 1990) and FASTA (Pearson, 1990) use more efficient heuristic-based alignment algorithms. BLOSUM (Altschul, et al., 1990) and PAM (Henikoff and Henikoff, 1992) are two widely-used amino acid substitution matrices to score similarity of two aligned residues. An affine function is used to penalize gaps (i.e., unaligned residues) in an alignment. Generally speaking, sequence-based alignment only works well for the alignments of close homologous proteins (sequence identity > 40%) since there are many conserved residues in their alignment and few gaps. The limitation of sequence alignment lies in that it cannot reliably make alignments when proteins under study are not very close to each other especially when the similarity of two proteins falls into the twilight zone, i.e., the sequence identity of two proteins is less than 25%. However, in many cases two proteins sharing low sequence identity may still be homologous and share some important structural and functional properties.

The quality of alignment can be improved by using sequence profile. Sequence profile is built on the Multiple Sequence Alignment (MSA) with sequence homologs, carries extra evolutionary information than its amino acid sequence alone. The intuition is that mutation frequency is position-specific and it can be detected when enough close homologs are found. To build a sequence profile, PSI-BLAST can be used to find close homologs of target protein from a large sequence database such as the NCBI non-redundant (NR) database and then build a MSA of these homologs and convert the MSA to a sequence profile. Various methods have been developed to align one primary sequence to one sequence profile or align two sequence profiles together. HMMER (Eddy, 2001) and SAM (Hughey and Krogh, 1995) are two tools that align one primary sequence to one profile HMM. Other sequence-profile alignment tools include DIALIGN (Morgenstern, et al., 1998) and FFAS (Jaroszewski, et al., 2005). HHpred (Söding, 2005), FORTE (Tomii and Akiyama, 2004), and PICASSO (Heger and Holm, 2001) are some tools that use profile-profile alignment. They have shown better performance than sequence-sequence or sequence-profile methods. PSI-BLAST (Altschul, et al., 1997) can be used to generate sequence profile of a protein.

The quality of profile-based alignment also depends on the representation of sequence profile. PSI-BLAST represents the sequence profile as a position-specific frequency matrix (PSFM) or position-specific scoring matrix (PSSM), which is widely-used in many applications such as homology detection, fold recognition and protein structure prediction. Both PSFM and PSSM have dimension of 20×$N$, where $N$ is the protein sequence length. Each column in a PSFM contains the occurring frequency of 20 amino acids at the corresponding sequence position. Accordingly, each column in a PSSM contains the potential of mutating to 20 amino acids at the corresponding position. A good sequence profile shall include as much information in the MSA as possible. In addition to representation, the quality of a sequence profile depends on the following factors: the number of PSI-BLAST iterations, the E-value cutoff used to determine if two proteins are homologous or not, and the sequence weighting scheme. It also depends on how to include amino acid pseudo-counts in converting amino acid occurring frequency to mutation potential.

Profile HMM (Hidden Markov Model) is another way to model an MSA of protein homologs. Profile HMM is better than PSFM/PSSM in that the former takes into consideration correlations between adjacent residues and also explicitly models gaps, so profile HMM on average is more sensitive than PSSM/PSFM for protein alignment and remote homology detection. In particular, a profile HMM usually contains three states: match, insert and delete. A 'match' state at an MSA column models the probability of residues being allowed in the column. It also contains emission probability of each amino acid type at this column. An 'insert' or 'delete' state at an MSA column allow for insertion of residues between that column and the next, or for deletion of residues. That is, a profile HMM has a position-dependent gap penalty. The penalty for an insertion or deletion depends on the HMM model parameters in each position. By contrast, traditional sequence alignment model uses a position-independent gap penalty. An insertion or deletion of $x$ residues is typically scored with an affine gap penalty, say $a + b(x-1)$ where $a$ is the penalty for a gap opening and $b$ for an extended gap.

Profile-based alignment method can fail when a protein has a very sparse sequence profile. The sparseness of a sequence profile can be quantified using the number of effective sequence homologs (NEFF). NEFF can also be interpreted as the average Shannon 'sequence entropy' for the profile or the average number of amino acid substitutions across all residues of a protein. The NEFF at one residue is calculated by $exp(-\sum_k p_k lnp_k)$ where $p_k$ is the probability for the *k*-th amino acid type, and the NEFF for the whole protein is the average across all residues. Therefore, NEFF ranges from 1 to 20 (i.e. the number of amino acid types). A smaller NEFF corresponds to a sparser sequence profile and less homologous information content. To go beyond this limitation, structural information can be incorporated in building alignments. For the target protein with unknown structure, we can predict its various structural information, such as 3 types and 8 types of secondary structure and solvent accessibility. If the predicted structural features are aligned approximately correct with the true secondary structure features of the template, the potential search space of possible alignments would be reduced substantially. Popular threading algorithms such as RAPTOR (Xu, et al., 2003), MUSTER (Wu and Zhang, 2008), HHpred (Söding, 2005) and Sparks (Yang, et al., 2011) all exploit the structural features to build alignments. Structural features are shown to be effective especially for proteins with sparse profile.

Another important component of alignment approach is its scoring function, which calculates a ratio between the likelihood of two proteins being homologous (or evolutionarily related) and that of being non-homologous (or evolutionarily unrelated). For sequence-sequence alignment, we can use two amino acid substitution models to estimate the probability of two proteins being homologous and non-homologous, respectively. The probability model for "non-homologous" is also called null model, describing the case that two aligned residues are evolutionarily unrelated. A few probability models such as PAM and BLOSUM have been developed to estimate how likely two aligned residues are evolutionarily related. PAM estimates the relatedness of two aligned residues starting from single point mutations. BLOSUM derives amino acid substitution model from blocks of multiple sequence alignment.

Unlike protein sequence alignment that uses an amino acid substitution matrix such as BLOSUM62, profile-based alignment needs a different scoring function. Nevertheless, some scoring functions for primary sequence-based homology detection can be generalized to profile-based methods. A slight change of the scoring functions can apply to the case when a profile is represented as an HMM. Given a primary sequence/profile and a profile from another sequence, to determine their similarity, one strategy is to estimate how likely the primary sequence is a sample from the probability distribution encoded by the sequence profile. The larger the alignment score the more likely that the primary sequence/profile is a sample from the probability distribution encoded by the sequence profile. Therefore, the alignment score quantifies the similarity between the primary sequence/profile and the sequence profile. Methods implemented such idea includes HMMER and HHpred. Other profile-based alignment scoring function such as dot product and Jensen-Shannon scores are also proposed in literature (Yona and Levitt, 2002).

# Chapter 3. Protein Alignment by Using Markov Random Fields

## 3.1 Introduction

As mentioned in previous chapter, all the popular profile-based alignment methods represent an MSA with a position-specific scoring matrix (PSSM) or an HMM (Hidden Markov Model). In this chapter, I will describe a Markov Random Field (MRF) representation of sequence profiles. That is, an MRF is used to model a multiple sequence alignment (MSA) of close sequence homologs. Compared to HMM that can only model local-range residue correlation, MRF can model long-range residue interactions (e.g., residue co-evolution) and thus, encodes global information in a protein. An MRF is a graphical model encoding a probability distribution over the MSA by a graph and a set of preset statistical functions. A node in the MRF corresponds to one column in the MSA and the existence of an edge between two nodes specifies correlation between two columns. Each node is associated with a function describing position-specific amino acid mutation pattern. Similarly, each edge is associated with a function describing correlated mutation statistics between two columns. Using MRF to represent the profiles, alignment of two proteins or protein families becomes that of two MRFs. To align two MRFs, a scoring function or alignment potential is needed to measure the similarity of two MRFs. We use a scoring function that consists of both node alignment potentials and edge alignment potential, which measure the node (i.e., amino acid) similarity and edge (i.e., interaction pattern) similarity, respectively. We will introduce the two scoring functions in the next chapters.

The graph of MRF derived profile MSA might contain loops so it is computationally challenging to optimize a scoring function containing edge alignment potential. To deal with this, we formulate MRF-MRF alignment as an Integer Linear Programming (ILP) problem and then develop an ADMM (Alternative Direction Method of Multipliers) (Boyd, et al., 2011) algorithm to dentify an approximate (sub-optimal) solution. ADMM divides the MRF alignment problem into two tractable sub-problems and then iteratively solve them until they converge. Experiments show that this MRF-MRF alignment method, denoted as MRFalign (Ma, et al., 2014), can generate more accurate alignments and is also much more sensitive than other methods. MRFalign works particularly well on mainly-beta proteins.

## 3.2 Methods

**Modeling a protein family using Markov Random Fields**

Given a protein sequence, we run PSI-BLAST with 5 iterations and E-value cutoff 0.001 to find its sequence homologs and then build their MSA (multiple sequence alignment). We can use a multivariate random variable $X = (X_1, X_2, ..., X_N)$, where $N$ is the number of columns (or the MSA length), to model the MSA. As shown in Figure 3, each $X_i$ is a finite discrete random variable representing the amino acid at column i in the MSA, taking values from 1 to 21, corresponding to 20 amino acids and gap. The occurring probability of the whole MSA can be modeled by MRF that is a function of $X$. An MRF node represents one column in the MSA and an edge represents the correlation between two columns. Here we ignore very short-range residue correlation since it is not very informative. An MRF consists of two types of functions: $\phi(X_i)$ and $\psi(X_i, X_k)$, where $\phi(X_i)$ is an amino acid preference function for node $i$ and $\psi(X_i, X_k)$ is a pairwise amino acid preference function for edge $(i, k)$ that reflects interaction between two nodes. Then, the probability of observing a particular protein sequence $X$ can be calculated as follows.

$$P(X) = \frac{1}{Z} \prod_i \phi(X_i) \prod_{(i,k)} \psi(X_i, X_k) \tag{1}$$

where $Z$ is the normalization factor (i.e., partition function).

The potential functions takes two kinds of information as features. One is the occurring probability of 20 amino acids and gap at each node (i.e., each column in MSA), which can also be interpreted as the marginal probability at each node. The other is the correlation between two nodes, which can be interpreted as interaction strength of two MSA columns.

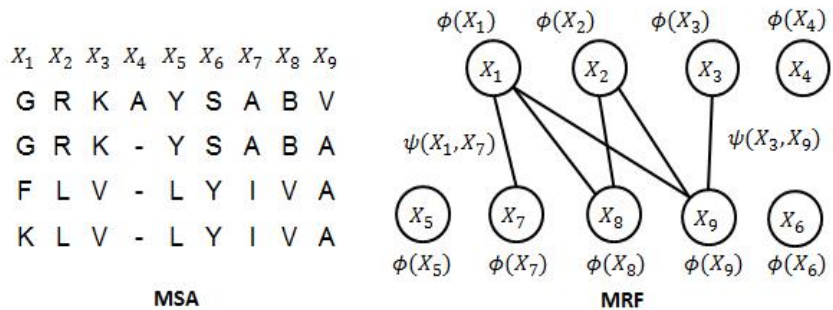

**Figure 3.** Modeling a multiple sequence alignment (left) by a Markov Random Field (right).

**Scoring function for the alignment of two Markov Random Fields (MRFs)**

Our scoring function for MRF-MRF alignment is a linear combination of node alignment potential and edge alignment potential with equal weight. Let $T$ and $S$ denote two MRFs for the two proteins under consideration. There are three possible alignment states $M$, $I_t$ and $I_s$ where $M$ represents two nodes being aligned, $I_t$ denotes an insertion in $T$ (i.e., one node in $T$ is not aligned), and $I_s$ denotes an insertion in $S$ (i.e., one node in $S$ is not aligned). As shown in Figure 4, each alignment can be represented as a path in an alignment matrix, in which each vertex can be exactly determined by its position in the matrix and its state. For example, the first vertex in the path can be written as $(0, 0, dummy)$, the 2nd vertex as $(1, 1, M)$ and the 3rd vertex as $(1, 1, I_s)$. Therefore, we can write an alignment as a set of triples, each of which has a form like $(i, j, u)$ where $(i, j)$ represents the position and $u$ the state.

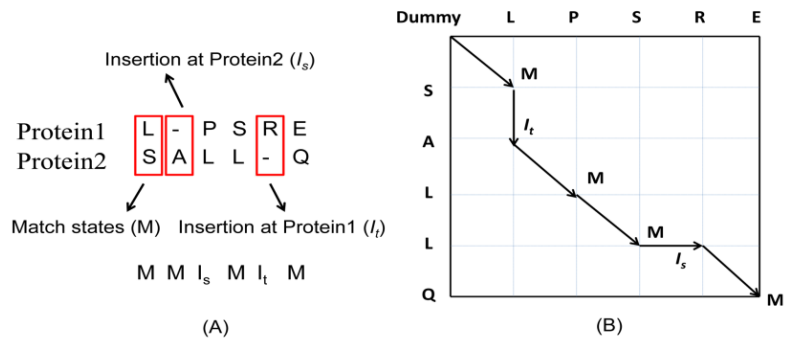

**Figure 4.** Representation of protein alignment. (A) Represented as a sequence of states. (B) Each alignment is a path in the alignment matrix.

**Scoring similarity of two Markov Random Fields**

This section will introduce how to align two proteins by aligning their corresponding MRFs. As shown in the left picture of Figure 5, building an alignment is equivalent to finding a unique path from the left-top corner to the right-bottom corner of the alignment matrix. For each vertex along the path, we need a score to measure how good it is to transit to the next vertex. That is, we need to measure how similar two nodes of the two MRFs are. We call this kind of scoring function node alignment potential. Second, in addition to measuring the similarity between the two aligned MRF nodes, we want to quantify the similarity between two MRF edges. For example, in the right picture of Figure 5 residues "$L$" and "$S$" of the first protein are aligned to residues "$A$" and "$Q$" of the 2nd protein, respectively. We would like to estimate how good it is to align the pair $(L, S)$ to the pair $(A, Q)$. This pairwise similarity function is a function of two MRF edges and we call it the edge alignment potential. When the edge alignment potential is used to score the similarity of two MRFs, Viterbi algorithm or other simple dynamic programming cannot be used to find the optimal alignment. It can be proved that when edge alignment potential is considered and gaps are allowed, the MRF-MRF alignment problem is NP hard (Lathrop, 1994). In this work, we will describe an ADMM algorithm to quickly find a suboptimal alignment of two MRFs. Although suboptimal, we have empirically found that the resulting alignments exhibit high accuracies.

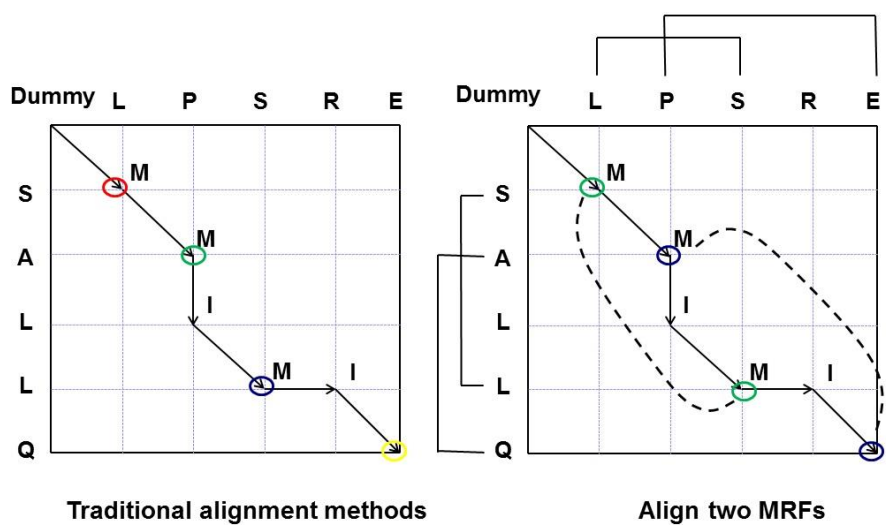

**Figure. 5.** Traditional alignment methods (left) and our MRFalign method (right)

## Node alignment potential of Markov Random Fields

Given an alignment, its node alignment potential is the accumulative potential of all the vertices in the path. We use a Conditional Neural Fields (CNF) (Ma, et al., 2012; Peng, et al., 2009) method to estimate the occurring probability of an alignment, and then derive node alignment potential from this CNF. Briefly speaking, we estimate the probability of an alignment A between $T$ and $S$ as follows.

$$P(A|T,S) = e^{\sum_{(i,j,u) \in A} E_u(T_i, S_j)} / Z(T, S) \quad (2)$$

where $Z(T, S)$ is a normalization factor summarizing all the possible alignments between T and S, and $E_u(T_i, S_j)$ is a neural network with one hidden layer that calculates the log-likelihood of a vertex $(i, j, u)$ in the alignment path, where i is a node in $T$, $j$ a node in $S$, and $u$ a state. When $u$ is a match state, $E_u$ takes as input the sequence profile context of two nodes $i$ and $j$, denoted as $T_i$ and $S_j$, respectively, and yields the log-likelihood of these two nodes being matched. When $u$ is an insertion state, it takes as input the sequence profile context of one node and yields the log-likelihood of this node being an insertion. The sequence profile context of node $i$ is a $21 \times (2w + 1)$ matrix where $w = 5$, consisting of the marginal probability of 20 amino acids and gap at $2w + 1$ nodes indexed by $i - w$, $i - w + 1, ..., i, i + 1, ..., i + w$. In case that one column does not exist (when $i \leq w$ or $i + w > N$), zero is used. We train the parameters in $E_u$ by maximizing the occurring probability of a set of reference alignments, which are generated by a structure alignment tool DeepAlign (Wang, et al., 2013). That is, we optimize the model parameters so that the structure alignment of one training protein pair has the largest probability among all possible alignments. A $L_2$-norm regularization factor, which is determined by 5-fold cross validation, is used to restrict the search space of model parameters to avoid over-fitting. See chapter 4 for more technical details.

Let $\theta_{i,j}^u$ denote the local alignment potential of a vertex $(i, j, u)$ in the alignment path. We calculate $\theta_{i,j}^u$ from $E_u$ as follows.

$$\theta_{i,j}^u = E_u(T_i, S_j) - Exp(E_u) \quad (3)$$

where $Exp(E_u)$ is the expected value of $E_u$. It is used to offset the effect of the background, which is the log-likelihood yielded by $E_u$ for any randomly chosen node pairs (or nodes). We can calculate the reference alignment likelihood $E_u$ in Eq. (3) by randomly sampling a set of protein pairs, each with the same lengths as the sequence $S$ and template $T$, respectively, and then estimating the probability of alignment A based upon these randomly sampled protein pairs. As long as we generate a sufficient number of samples, we can accurately approximate $E_u$. Here, $E_u$ depends only on the alignment state but not any specific protein pair. I will introduce more details including the parameter estimation and background probability calculation in next few chapters.

**Edge alignment potential of Markov Random Fields.**

The edge alignment potential calculates the similarity of two edges, one from each MRF, based upon the interaction strength of two ends in one edge as shown in Figure 6. As mentioned above, we use a predicted distance probability distribution based on the features of two nodes to estimate their interaction strength. Let $d_{ik}^T$ be a random variable for the Euclidean distance between two residues at $i$ and $k$ and $d_{jl}^S$ is defined similarly. Let $\theta_{i,k,j,l}$ denote the alignment potential between edge $(i,k)$ in T and edge $(j,l)$ in S. We can calculate $\theta_{i,k,j,l}$ as follows.

$$\theta_{i,k,j,l} = \sum_{d_{ik}^T, d_{jl}^S} p(d_{ik}^T|c_i, c_k, m_{ik}) p(d_{jl}^S|c_j, c_l, m_{jl}) \log \frac{p(d_{ik}^T, d_{jl}^S)}{P_{ref}(d_{ik}^T) P_{ref}(d_{jl}^S)} \quad (4)$$

where $p(d_{ik}^T|c_i, c_k, m_{ik})$ is the probability of two nodes i and k in $T$ interacting at distance $d_{ik}^T$; $p(d_{jl}^S|c_j, c_l, m_{jl})$ is the probability of two nodes $j$ and $l$ in $S$ interacting at distance $d_{jl}^S$; $c_i$ and $c_k$ are the sequence profile contexts of two nodes $i$ and $k$, respectively, and $m_{ik}$ represents the Mutual Information (MI) (or interaction strength) between these two nodes. The sequence profile context of node i is a $21 \times (2w+1)$ matrix where $w = 5$, consisting of the occurring probability of 20 amino acids and gap at $2w+1$ nodes indexed by $i-w, i-w+1, \ldots, i, i+1, \ldots, i+w$. In case that one column does not exist (when $i \leq w$ or $i + w > N$), zero is used. $p(d_{ik}^T, d_{jl}^S)$ is the probability of one distance $d_{ik}^T$ being aligned to another distance $d_{jl}^S$ in reference alignments; and $P_{ref}(d_{ik}^T)$ ($P_{ref}(d_{jl}^S)$) is the background probability of observing $d_{ik}^T$ ($d_{jl}^S$) in a protein structure. Meanwhile $x_i$ and $x_k$ are position-specific features centered at the *i*th and *k*th residues, respectively, and $m_{ik}$ represents the mutual information between the *i*th and *k*th columns in the multiple sequence alignment. We predict $p(d_{ik}^T|c_i, c_k, m_{ik})$ using a probabilistic neural network (PNN) implemented in our context-specific distance-dependent statistical potential package EPAD (Zhao and Xu, 2012). EPAD takes as input sequence contexts and co-evolution information and then yields inter-residue distance probability distribution. Compared to contact information, here we use interaction at a given distance to obtain a higher-resolution description of the residue interaction pattern.

Therefore, our scoring function contains more information and thus, may yield better alignment accuracy.

Instead of using power series of MI, we can use Direct Information (DI), which is a global statistics (i.e., measuring the residue co-evolution strength of two positions considering other positions). DI can be calculated by some contact prediction programs such PSICOV (Jones, et al., 2012), Evfold (Marks, et al., 2011), plmDCA (Ekeberg, et al., 2013) as residue co-evolution. PSICOV assumes that $P(X)$ is a Gaussian distribution and calculates the partial correlation between two columns by inverse covariance matrix. By contrast, plmDCA does not assume a Gaussian distribution and is more efficient and also slightly more accurate. Generally speaking, these programs are time-consuming. The reliability of mutual information (MI) or direct information (DI) depends on the number of non-redundant sequence homologs. When there are few sequence homologs, the resulting MI or DI is not very accurate. Therefore, it is not enough to only use residue co-evolution strength to estimate residue interaction strength. We can use other contact prediction programs such as PhyCMAP (Wang and Xu, 2013) which integrates both residue co-evolution information, PSI-BLAST sequence profile and others to predict the probability of two residues in contact. In Chapter 5 I will introduce a new computational approach to estimate the interaction strength between residues integrating joint multi-family evolutionary coupling analysis and supervised learning.

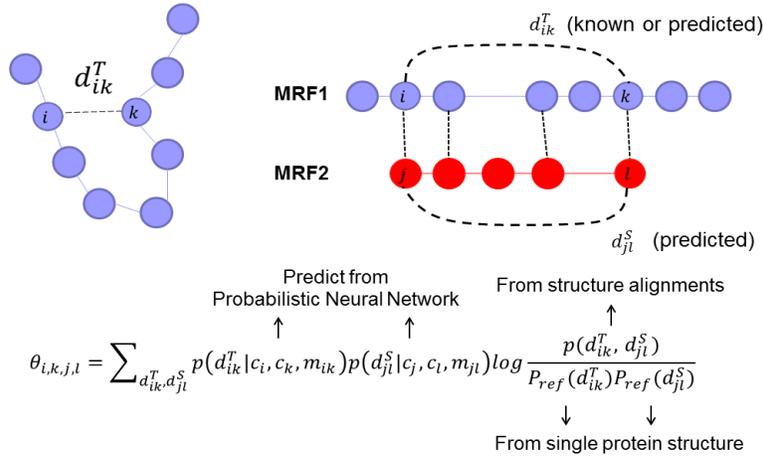

**Figure 6.** Illustration of edge alignment potential for MRF-MRF alignment.

**Aligning two MRFs by ADMM (Alternating Direction Method of Multipliers)**

As mentioned before, an alignment can be represented as a path in the alignment matrix, which encodes an exponential number of paths. We can use a set of $3N_1N_2$ binary variables to indicate which path is chosen, where $N_1$ and $N_2$ are the lengths of the two MSAs, $(i, j)$ is an entry in the alignment matrix and $u$ is the associated state. $z_{i,j}^u$ is equal to 1 if the alignment path passes $(i, j)$ with state $u$. Therefore, the problem of finding the best alignment between two MRFs can be formulated as the following quadratic optimization problem.

$$(P1) \quad max_z \sum_{i,j,u} \theta_{i,j}^u z_{i,j}^u + \frac{1}{L}\sum_{i,j,k,l,u,v} \theta_{i,j,k,l}^{uv} z_{i,j}^u z_{k,l}^v \quad (5)$$

where $\theta_{i,j}^u$ and $\theta_{i,j,k,l}^{uv}$ are node and edge alignment potentials as described in previous section. Meanwhile, $\theta_{i,j,k,l}^{uv}$ is equal to 0 if either $u$ or $v$ is not a match state. $L$ is the alignment length and $1/L$ is used to make the accumulative node and edge potential have similar scale. Note that $L$ is unknown and we will describe how to determine it later in this section. Finally, the solution of P1 shall be subject to the constraint that all those $z_{i,j}^u$ with value 1 shall form a valid alignment path. This constraint shall also be enforced to all the optimization problems described in this section.

It is computationally intractable to find the optimal solution of P1. Below we present an ADMM (Alternating Direction Method of Multipliers) method that can efficiently solve this problem to suboptimal. See (Boyd, et al., 2011) for a tutorial of the ADMM method. To use ADMM, we rewrite P1 as follows by making a copy of $z$ to $y$, but without changing the solution space.

$$(P2) \quad max_{z,y} \sum_{i,j,u} \theta_{i,j}^u z_{i,j}^u + \frac{1}{L}\sum_{i,j,k,l,u,v} \theta_{i,j,k,l}^{uv} z_{i,j}^u y_{k,l}^v \quad (6)$$

$$s.t. \quad \forall k, l, v, \quad z_{k,l}^v = y_{k,l}^v$$

Problem P2 can be augmented by adding a term to penalize the difference between $z$ and $y$.

$$(P3) \quad max_{z,y} \sum_{i,j,u} \theta_{i,j}^u z_{i,j}^u + \frac{1}{L}\sum_{i,j,k,l,u,v} \theta_{i,j,k,l}^{uv} z_{i,j}^u y_{k,l}^v - \frac{\rho}{2}\sum_{i,j,u}(z_{i,j}^u - y_{i,j}^u)^2 \quad (7)$$

$$s.t. \quad \forall i, j, u, \quad z_{i,j}^u = y_{i,j}^u$$

P3 is equivalent to P2 and P1, but converges faster due to the penalty term. Here $\rho$ is a hyper-parameter influencing the convergence rate of the algorithm. Empirically, setting $\rho$ to a constant (=0.5) enables our algorithm to converge within 10 iterations for most protein pairs.

Adding the constraint $z_{i,j}^u = y_{i,j}^u$ using a Lagrange multiplier $\lambda$ to Eq. (7), we have the following Lagrangian dual problem:

$$(P4) \quad min_\lambda max_{z,y} \sum_{i,j,u} \theta_{i,j}^u z_{i,j}^u + \frac{1}{L}\sum_{i,j,k,l,u,v} \theta_{i,j,k,l}^{uv} z_{i,j}^u y_{k,l}^v + \sum_{i,j,u} \lambda_{i,j}^u (z_{i,j}^u - y_{i,j}^u) - \frac{\rho}{2}\sum_{i,j,u}(z_{i,j}^u - y_{i,j}^u)^2 \quad (8)$$

It is easy to prove that P3 is upper bounded by P4. Now we will solve P4 and use its solution to approximate P3 and thus, P1. Since both $z$ and $y$ are binary variables, the last term in Eq. (8) can be expanded as follows.

$$\frac{\rho}{2}\sum_{i,j,u}(z_{i,j}^u - y_{i,j}^u)^2 = \frac{\rho}{2}\sum_{i,j,u}(z_{i,j}^u + y_{i,j}^u - 2z_{i,j}^u y_{i,j}^u) \tag{9}$$

For a fixed $\lambda$, we can split P4 into the following two sub-problems.

$$\text{(SP1)} \qquad y^* = argmax \sum_{k,l,v} y_{k,l}^v C_{k,l}^v \tag{10}$$

$$\text{where } C_{k,l}^v = \frac{1}{L}\sum_{i,j,u} \theta_{i,j,k,l}^{uv} z_{i,j}^u - \lambda_{k,l}^v - \frac{\rho}{2}(1 - 2z_{k,l}^v)$$

$$\text{(SP2)} \qquad z^* = argmax \sum_{i,j,u} z_{i,j}^u D_{i,j}^u \tag{11}$$

$$\text{where } D_{i,j}^u = \theta_{i,j}^u + \sum_{k,l,v} \frac{1}{L}\theta_{i,j,k,l}^{uv} y_{k,l}^{v*} + \lambda_{i,j}^u - \frac{\rho}{2}(1 - y_{i,j}^{u*})$$

The sub-problem SP1 optimizes the objective function with respect to $y$ while fixing $z$, and the sub-problem SP2 optimizes the objective function with respect to $z$ while fixing $y$. SP1 and SP2 do not contain any quadratic term, so they can be efficiently solved using the classical dynamic programming algorithm for sequence or HMM-HMM alignment.

In summary, we solve P4 using the following procedure,

**Step 1:** Initialize $z$ by aligning the two MRFs without the edge alignment potential, which can be done by dynamic programming. Accordingly, initialize $L$ as the length of the initial alignment.

**Step 2:** Solve (SP1) first and then (SP2) using dynamic programming, each generating a feasible alignment.

**Step 3:** If the algorithm converges, i.e., the difference between $z$ and $y$ is very small or zero, stop here. Otherwise, we update the alignment length $L$ as the length of the alignment just generated and the Lagrange multiplier $\lambda$ using sub-gradient descent as in Eq. (12), and then go back to Step 2.

$$\lambda^{n+1} = \lambda^n - \rho(z^* - y^*) \tag{12}$$

Due to the quadratic penalty term in Eq. (6), this ADMM algorithm usually converges much faster and also yields better solutions than without this term. Empiric ally, it converges within 10 iterations for most protein pairs. See (Boyd, et al., 2011) for the convergence proof of a general ADMM algorithm.

## 3.3 Results

**Training and validation data**

To train the node alignment potential, we constructed the training and validation data from SCOP70. The sequence identity of all the training and validation protein pairs is uniformly distributed between 20% and 70%. Further, two proteins in any pair are similar at superfamily or fold level. In total we use a set of 1400 protein pairs as the training and validation data, which covers 458 SCOP folds (Andreeva, et al., 2004). Five-fold cross validation is used to choose the hyper-parameter in our machine learning model. In particular, every time we choose 1000 out of the 1400 protein pairs as the training data and the remaining 400 pairs as the validation data such that there is no fold-level redundancy between the training and validation data. A training or validation protein has less than 400 residues and contains less than 10% of residues without 3D coordinates. The reference alignment for a protein pair is generated by a structure alignment tool DeepAlign. Each reference alignment has fewer than 50 gap positions in the middle and the number of terminal gaps is less than 20% of the alignment length.

**Test data**

The data used to test alignment accuracy has no fold-level overlap with the training and validation data. In particular, we use the following three datasets to test the alignment accuracy, which are subsets of the test data used in (Angermüller, et al., 2012) to benchmark protein modeling methods.

1. Set3.6K: a set of 3617 non-redundant protein pairs. Two proteins in a pair share <40% sequence identity and have small length difference. By "non-redundant" we mean that in any two protein pairs, there are at least two proteins (one from each pair) sharing less than 25% sequence identity.

2. Set2.6K: a set of 2633 non-redundant protein pairs. Two proteins in a pair share <25% sequence identity and have length difference larger than 30%. This set is mainly used to test the performance of one method in handling with domain boundary.

3. Set60K: a very large set of 60929 protein pairs, in most of which two proteins share less than 40% sequence identity. Meanwhile, 846, 40902, and 19181 pairs are similar at the SCOP family, superfamily and fold level, respectively, and 151, 2691 and 2218 pairs consist of only all-beta proteins, respectively.

We use the following benchmarks to test remote homology detection success rate.

4. SCOP20, SCOP40 and SCOP80, which are used by Söding group to study context-specific mutation score (Angermüller, et al., 2012). They are constructed by filtering the SCOP database with a maximum sequence identity of 20%, 40% and 80%, respectively. In total they have 4884, 7088, and 9867 proteins, respectively, and 1281, 1806, and 2734 beta proteins, respectively.

We run PSI-BLAST with 5 iterations to detect sequence homologs and generate MSAs for the first three datasets. The MSA files for the three SCOP benchmarks are downloaded from the HHpred website. Pseudo-counts are used in building sequence profiles. Real secondary structure information is not used since this paper focuses on sequence-based homology detection.

**Programs to compare**

To evaluate alignment accuracy, we compare our method, denoted as MRFalign, with sequence-HMM alignment method HMMER and HMM-HMM alignment method HHalign. HHMER is run with a default E-value threshold (10.0). HHalign is run with the option "-mact 0.1". To evaluate the performance of homology detection, we compare MRFalign, with FFAS (Jaroszewski, et al., 2005) (PSSM-PSSM comparison), hmmscan (sequence-HMM comparison) and HHsearch and HHblits (Remmert, et al., 2012) (HMM-HMM comparison). HHsearch and hmmscan use HHalign and HMMER, respectively, for protein alignment.

**Evaluation criteria**

Three performance metrics are used including reference-dependent alignment precision, alignment recall and homology detection success rate. Alignment precision is defined as the fraction of aligned positions that are correctly aligned. Alignment recall is the fraction of alignable residues that are correctly aligned. Reference alignments are used to judge if one residue is correctly aligned or alignable. To reduce bias, we use three very different structure alignment tools to generate reference alignments, including TM-align (Zhang and Skolnick, 2005), Matt (Menke, et al., 2008), and DeepAlign.

**Reference-dependent alignment recall**

As shown in Tables 1 and 2, our method MRFalign exceeds all the others regardless of the reference alignments on both dataset Set3.6K and Set2.6K.

MRFalign outperforms HHalign by ~10% on both datasets, and HHMER by ~23% and ~24%, respectively. If 4-position off the exact match is allowed in calculating alignment recall, MRFalign outperforms HHalign by ~11% on both datasets, and HHMER by ~25% and ~33%, respectively.

**Table 1.** Reference-dependent alignment recall on Set3.6K. Three structure alignment tools (TMalign, Matt and DeepAlign) are used to generate reference alignments. "4-offset" means that 4-position off the exact match is allowed. The bold indicates the best results.

|  | TMalign | | Matt | | DeepAlign | |
|---|---|---|---|---|---|---|
|  | Exact match | 4-offset | Exact match | 4-offset | Exact match | 4-offset |
| HMMER | 22.9% | 26.5% | 24.1% | 27.4% | 25.5% | 28.1% |
| HHalign | 36.3% | 39.1% | 37.0% | 42.1% | 38.4% | 42.8% |
| MRFalign | **47.4%** | **51.0%** | **47.5%** | **52.6%** | **49.2%** | **53.5%** |

**Table 2.** Reference-dependent alignment recall on Set2.6K. See Table 1 for explanation.

|  | TMalign | | Matt | | DeepAlign | |
|---|---|---|---|---|---|---|
|  | Exact match | 4-offset | Exact match | 4-offset | Exact match | 4-offset |
| HMMER | 36.5% | 42.6% | 38.6% | 44.0% | 40.4% | 45.0% |
| HHalign | 62.5% | 66.1% | 63.2% | 66.2% | 64.0% | 66.7% |
| MRFalign | **72.8%** | **76.2%** | **73.5%** | **76.7%** | **74.2%** | **77.8%** |

On the very large set Set60K, as shown in Table 3, our method outperforms the other two in each SCOP classification regardless of the reference alignments used. MRFalign is only slightly better than HHalign at the family level, which is not surprising since it is easy to align two closely-related proteins. At the superfamily level, our method outperforms HHalign and HMMER by ~6% and ~18%, respectively. At the fold level, our method outperforms HHalign and HHMER by ~7% and ~14%, respectively.

**Alignment recall for beta proteins.** Our method outperforms HHalign and HMMER by ~3% and ~12%, respectively, at the family level; ~7% and ~19%, respectively, at the superfamily level; and ~10% and ~16%, respectively, at the fold level, regardless of reference alignments.

**Table 3.** Reference-dependent alignment recall (exact match) on the large benchmark Set60K. The protein pairs are divided into 3 groups based upon the SCOP classification. The bold indicates the best results.

|  | TMalign | Matt | DeepAlign |
|---|---|---|---|

|  | HMMER | HHalign | MRFalign | HMMER | HHalign | MRFalign | HMMER | HHalign | MRFalign |
|---|---|---|---|---|---|---|---|---|---|
| Family | 57.4% | 69.2% | **71.0%** | 59.1% | 70.5% | **74.5%** | 63.2% | 72.6% | **75.5%** |
| Superfamily | 31.2% | 42.0% | **48.1%** | 32.3% | 42.4% | **51.7%** | 32.8% | 49.4% | **55.6%** |
| Fold | 1.3% | 7.0% | **14.2%** | 1.6% | 8.0% | **15.5%** | 2.0% | 8.7% | **18.4%** |
| Family (beta) | 60.9% | 69.9% | **73.1%** | 64.0% | 75.1% | **78.4%** | 68.4% | 79.0% | **82.9%** |
| Superfamily (beta) | 35.0% | 47.2% | **52.1%** | 37.0% | 50.2% | **55.8%** | 39.1% | 52.9% | **60.7%** |
| Fold (beta) | 2.5% | 8.3% | **17.3%** | 3.0% | 9.1% | **17.1%** | 4.0% | 10.1% | **21.8%** |

**Reference-dependent alignment precision**

As shown in Tables 4 and 5, our method MRFalign exceeds all the others regardless of the reference alignments on both data sets Set3.6K and Set2.6K. MRFalign outperforms HHalign by ~8% and ~5%, respectively, and HMMER by ~15% and ~13%, respectively. If 4-position off the exact match is allowed in calculating alignment precision, MRFalign outperforms HHalign by ~8% and ~9%, and HMMER by ~14% and ~18% on Set3.6K and Set2.6K, respectively.

**Table 4.** Reference-dependent alignment precision on Se3.6K. Three structure alignment tools (TMalign, Matt and DeepAlign) are used to generate reference alignments. "4-offset" means that 4-position off the exact match is allowed. The bold indicates the best results.

|  | TMalign | | Matt | | DeepAlign | |
|---|---|---|---|---|---|---|
|  | Exact match | 4-offset | Exact match | 4-offset | Exact match | 4-offset |
| HMMER | 29.3% | 34.1% | 29.6% | 34.7% | 31.5% | 35.6% |
| HHalign | 35.9% | 39.4% | 36.2% | 39.4% | 37.2% | 41.7% |
| MRFalign | **43.2%** | **47.4%** | **44.1%** | **48.5%** | **46.1%** | **50.4%** |

**Table 5.** Reference-dependent alignment precision on Set2.6K. See Table 4 for explanation.

|  | TMalign | | Matt | | DeepAlign | |
|---|---|---|---|---|---|---|
|  | Exact match | 4-offset | Exact match | 4-offset | Exact match | 4-offset |
| HMMER | 48.0% | 50.1% | 48.2% | 50.3% | 51.4% | 54.8% |
| HHalign | 57.1% | 59.9% | 57.3% | 60.0% | 58.3% | 61.4% |
| MRFalign | **62.5%** | **69.1%** | **62.7%** | **69.6%** | **63.2%** | **70.0%** |

On the very large set Set60K, as shown in Table 6, our method outperforms the other two in each SCOP classification regardless of the reference alignments used. At the family level, our method outperforms HHalign and HMMER by ~3% and ~4%, respectively. At the superfamily level, our method outperforms HHalign and HMMER by ~4% and ~5%, respectively. At the fold level, our method outperforms HHalign and HMMER by ~5% and ~8%, respectively.

**Table 6.** Reference-dependent alignment precision (exact match) on the large benchmark Set60K. The protein pairs are divided into 3 groups based upon the SCOP classification. The bold indicates the best results.

|  | TMalign | | | Matt | | | DeepAlign | | |
|---|---|---|---|---|---|---|---|---|---|
|  | HMMER | HHalign | MRFalign | HMMER | HHalign | MRFalign | HMMER | HHalign | MRFalign |
| Family | 63.1% | 63.9% | **67.3%** | 64.3% | 65.4% | **68.0%** | 68.4% | 69.2% | **71.4%** |
| Superfamily | 38.7% | 39.5% | **42.8%** | 40.5% | 41.3% | **44.9%** | 43.2% | 44.3% | **48.7%** |
| Fold | 4.2% | 7.4% | **11.5%** | 4.7% | 8.0% | **12.3%** | 5.4% | 8.2% | **14.5%** |
| Family (beta) | 66.4% | 65.8% | **69.5%** | 67.4% | 68.1% | **72.3%** | 70.8% | 72.4% | **77.9%** |
| Superfamily (beta) | 44.2% | 44.9% | **48.8%** | 45.4% | 46.2% | **49.4%** | 46.6% | 48.4% | **53.7%** |
| Fold (beta) | 6.1% | 9.3% | **14.1%** | 6.7% | 9.2% | **14.5%** | 7.9% | 8.6% | **17.8%** |

**Homology detection success rate**

To evaluate homology detection rate, we employ three benchmarks SCOP20, SCOP40 and SCOP80 introduced in (Angermüller, et al., 2012). For each protein sequence in one benchmark, we treat it as a query, align it to all the other proteins in the same benchmark and then examine if those with the best alignment scores are similar to the query or not. We also conducted homology detection experiments using hmmscan, FFAS, HHsearch and HHblits with default options. The success rate is measured at the superfamily and fold levels, respectively. When evaluating the success rate at the superfamily (fold) level, we exclude those proteins similar to the query at least at the family (superfamily) level. For each query protein, we examine the top 1-, 5- and 10-ranked proteins, respectively.

As shown in Table 7, tested on SCOP20, SCOP40 and SCOP80 at the superfamily level, our method MRFalign succeeds on ~6%, ~4% and ~4% more query proteins than HHsearch, respectively, when only the first-ranked proteins are considered. As shown in Table 8, at the fold level, MRFalign succeeds on ~11%, ~11% and ~12% more proteins than HHsearch, respectively, when only the first-ranked proteins are evaluated. At the superfamily level, SCOP20 is more challenging than the other two benchmarks because it contains fewer proteins similar at this level. Nevertheless, at the fold level, SCOP80 is slightly more challenging than the other two benchmarks maybe because it contains many more irrelevant proteins and thus, the chance of ranking false positives at top is higher.

Similar to alignment accuracy, our method for homology detection also has a larger advantage on the beta proteins. In particular, as shown in Table 9, tested on SCOP20, SCOP40 and SCOP80 at the superfamily level, MRFalign succeeds on ~7%, ~5% and ~7% more proteins than HHsearch, respectively, when only the first-ranked proteins are evaluated. As shown in Table 10, at the fold level, MRFalign succeeds on ~13%, ~16% and ~17% more proteins than HHsearch, respectively, when only the first-ranked proteins are evaluated. Note that in this experiment, only the query proteins are mainly-beta proteins, the subject proteins can be of any types. If we restrict the subject proteins to only beta proteins, the success rate increases further due to the reduction of false positives.

**Table 7.** Homology detection performance at the superfamily level

|  | Scop20 | | | Scop40 | | | Scop80 | | |
|---|---|---|---|---|---|---|---|---|---|
|  | Top1 | Top5 | Top10 | Top1 | Top5 | Top10 | Top1 | Top5 | Top10 |
| hmmscan | 35.2% | 36.5% | 36.5% | 40.2% | 41.7% | 41.8% | 43.9% | 45.2% | 45.3% |
| FFAS | 48.6% | 54.4% | 55.6% | 52.1% | 56.3% | 57.1% | 49.8% | 53.0% | 53.7% |
| HHsearch | 51.6% | 57.3% | 59.2% | 55.8% | 60.8% | 62.4% | 56.1% | 60.1% | 61.8% |
| HHblits | 51.9% | 56.3% | 57.5% | 56.0% | 59.8% | 60.9% | 59.2% | 62.5% | 63.3% |
| MRFalign | **58.2%** | **61.7%** | **63.4%** | **59.3%** | **63.6%** | **65.8%** | **60.4%** | **64.7%** | **66.1%** |

**Table 8.** Homology detection performance at the fold level

|  | Scop20 | | | Scop40 | | | Scop80 | | |
|---|---|---|---|---|---|---|---|---|---|
|  | Top1 | Top5 | Top10 | Top1 | Top5 | Top10 | Top1 | Top5 | Top10 |
| hmmscan | 5.2% | 6.1% | 6.1% | 6.2% | 6.9% | 6.9% | 5.9% | 6.5% | 6.6% |
| FFAS | 13.1% | 18.7% | 20.0% | 10.4% | 14.5% | 15.4% | 9.1% | 11.9% | 12.6% |
| HHsearch | 16.3% | 24.7% | 28.6% | 17.6% | 25.3% | 29.1% | 15.4% | 21.9% | 25.0% |
| HHblits | 17.4% | 25.2% | 27.2% | 19.1% | 26.0% | 28.2% | 18.4% | 25.0% | 27.0% |
| MRFalign | **27.2%** | **36.8%** | **41.2%** | **28.3%** | **37.9%** | **42.4%** | **27.0%** | **38.1%** | **41.6%** |

**Table 9.** Homology detection performance for mainly beta proteins at the superfamily level

|  | Scop20 | | | Scop40 | | | Scop80 | | |
|---|---|---|---|---|---|---|---|---|---|
|  | Top1 | Top5 | Top10 | Top1 | Top5 | Top10 | Top1 | Top5 | Top10 |
| hmmscan | 29.1% | 29.4% | 29.4% | 34.7% | 35.1% | 35.1% | 43.7% | 44.0% | 44.1% |
| FFAS | 43.6% | 49.9% | 51.9% | 48.2% | 52.4% | 53.5% | 43.7% | 46.3% | 47.2% |
| HHsearch | 48.2% | 54.6% | 56.9% | 52.0% | 56.9% | 59.1% | 47.7% | 51.8% | 53.7% |
| HHblits | 47.5% | 52.1% | 53.7% | 51.4% | 54.8% | 56.6% | 52.9% | 54.6% | 57.8% |
| MRFalign | **55.4%** | **61.7%** | **65.9%** | **57.3%** | **63.5%** | **66.8%** | **54.2%** | **59.7%** | **64.2%** |

**Contribution of edge alignment potential and mutual information**

To evaluate the contribution of our edge alignment potential, we calculate the alignment recall improvement resulting from using edge alignment potential on two benchmarks Set3.6K and Set2.6K. As shown in Table 11, our edge alignment potential can improve alignment recall by 3.4% and 3.7%, respectively. When mutual information is used, we can further improve alignment recall by 1.1% and 1.9% on these two sets, respectively. Mutual information is mainly useful for proteins with many sequence homologs since it is close to 0 when there are few sequence homologs. As shown in Table 11, if only those proteins with at least 256 non-redundant sequence homologs are considered, the improvement resulting from mutual information is ~3%.

**Table 10.** Homology detection performance for mainly beta proteins at the fold level

|         | Scop20 | | | Scop40 | | | Scop80 | | |
|---------|--------|--------|--------|--------|--------|--------|--------|--------|--------|
|         | Top1 | Top5 | Top10 | Top1 | Top5 | Top10 | Top1 | Top5 | Top10 |
| hmmscan | 6.9% | 7.6% | 7.6% | 8.0% | 8.6% | 8.6% | 7.0% | 7.4% | 7.4% |
| FFAS | 22.7% | 30.1% | 31.8% | 15.2% | 20.4% | 21.7% | 11.8% | 15.3% | 16.1% |
| HHsearch | 24.4% | 34.7% | 38.8% | 26.8% | 37.7% | 41.6% | 19.1% | 26.8% | 29.5% |
| HHblits | 24.1% | 33.3% | 34.8% | 26.9% | 35.3% | 37.1% | 24.7% | 34.1% | 35.5% |
| MRFalign | **37.4%** | **55.0%** | **61.4%** | **42.5%** | **51.1%** | **54.6%** | **36.4%** | **48.0%** | **55.9%** |

**Table 11.** Contribution of edge alignment potential and mutual information, measured by alignment recall improvement on two benchmarks Set3.6K and Set2.6K. The structure alignments generated by DeepAlign are used as reference alignments.

| **Alignment recall for the whole test sets** | | | | |
|---|---|---|---|---|
| | Set3.6K | | Set2.6K | |
| | Exact Match | 4-position offset | Exact Match | 4-position offset |
| Only with node potential | 44.7% | 48.6% | 68.6% | 71.8% |
| Node + edge potential, no MI | 48.1% | 52.2% | 72.3% | 75.2% |
| Node + edge potential with MI | 49.2% | 53.5% | 74.2% | 77.8% |
| **Alignment recall on proteins with at least 256 non-redundant sequence homologs** | | | | |
| | 391 pairs in Set3.6K | | 509 pairs in Set2.6K | |
| Only with node potential | 59.5% | 63.4% | 71.3% | 75.8% |
| Node + edge potential, no MI | 62.1% | 66.7% | 73.5% | 78.1% |
| Node + edge potential with MI | 65.2% | 69.8% | 76.6% | 81.0% |

**Running time**

Figure 7 shows the running time of MRFalign with respect to protein length. As a control, we also show the running time of the Viterbi algorithm, which is used by our ADMM algorithm to generate alignment at each iteration. As shown in this figure, MRFalign is no more than 10 times slower than the Viterbi algorithm. To speed up homology detection, we first use the Viterbi algorithm to perform an initial search without considering edge alignment potential and keep only top 200 proteins, which are then subject to realignment and rerank by our MRFalign method. Therefore, although MRFalign may be very slow compared to the Viterbi algorithm, empirically we can do homology search only slightly slower than the Viterbi algorithm.

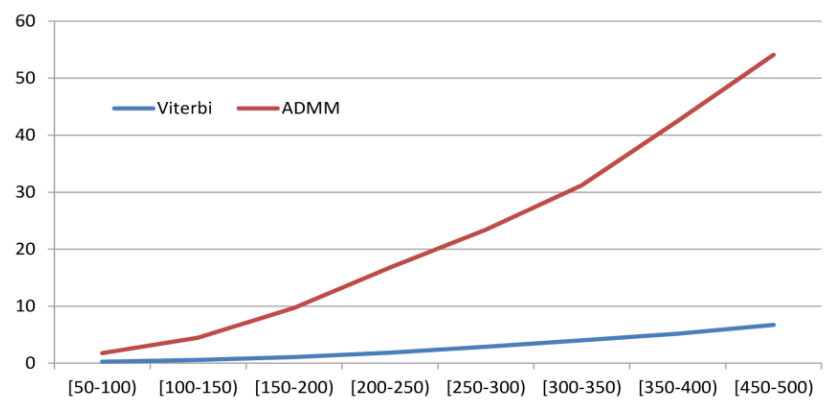

**Figure 7.** Running time of the Viterbi algorithm and our ADMM algorithm. The X-axis is the geometric mean of the two protein lengths in a protein pair. The Y-axis is the running time in seconds.

**Is our MRFalign method over-trained?**

We conducted two experiments to show that our MRFalign is not overtrained. In the first experiment, we used 36 CASP10 hard targets as the test data. Our training set was built before CASP10 started, so there is no redundancy between the CASP10 hard targets and our training data. Using MRFalign and HHpred, respectively, we search each of these 36 test targets against PDB25 to find the best match. Since PDB25 does not contain proteins very similar to many of the test targets, we built a 3D model using MODELLER from the alignment between a test target and its best match and then measure the quality of the model. As shown in Figure 8, MRFalign can yield much better 3D models than HHsearch for most of the targets. This implies that our method can generalize well to the test data not similar to the training data.

In the second experiment, we divide the proteins in SCOP40 into three subsets according their similarity with all the training data. We measure the similarity of one test protein with all the training data by its best BLAST E-value. We used two values 1e-2 and 1e-35 as the E-value cutoff so that the three subsets have roughly the same size. As shown in Table 12, the advantage of our method in remote homology detection over HHpred is roughly same across the three subsets. Since HHpred is an unsupervised algorithm, this implies that the performance of our method is not correlated to the test-training similarity. Therefore, it is unlikely that our method is overfit by the training data.

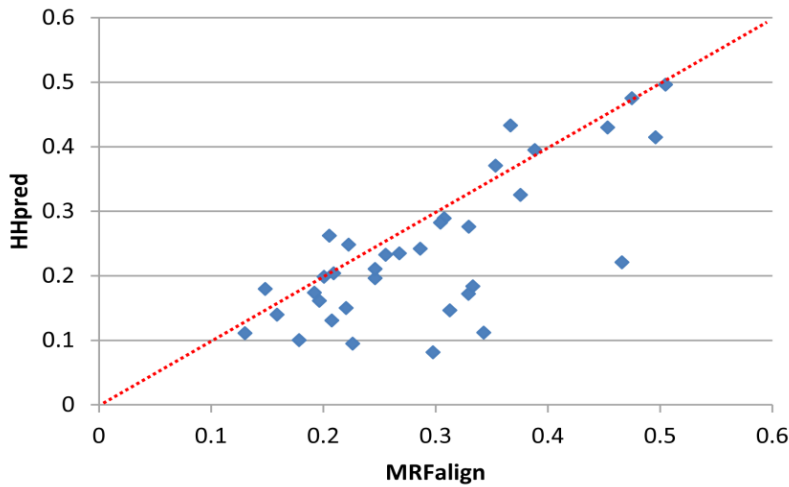

**Figure 8.** The model quality, measured by TM-score, of our method and HHpred for the 36 CASP10 hard targets. One point represents two models generated by our method (X-axis) and HHpred (Y-axis).

**Table 12.** Fold recognition rate of our method on SCOP40, with respect to the similarity (measured by E-value) between the test data and the training data.

|  | E-value < 1e-35 | | | 1e-35 < E-value < 1e-2 | | | E-value > 1e-2 | | |
| --- | --- | --- | --- | --- | --- | --- | --- | --- | --- |
|  | Top1 | Top5 | Top10 | Top1 | Top5 | Top10 | Top1 | Top5 | Top10 |
| hmmscan | 5.0% | 5.6% | 5.6% | 7.3% | 7.9% | 7.9% | 6.4% | 7.3% | 7.4% |
| FFAS | 10.3% | 14.5% | 15.8% | 9.7% | 12.9% | 13.5% | 11.6% | 16.5% | 17.5% |
| HHsearch | 16.0% | 23.2% | 26.5% | 18.5% | 26.2% | 30.3% | 18.9% | 27.2% | 31.7% |
| HHblits | 16.9% | 23.1% | 25.5% | 20.8% | 27.4% | 28.9% | 20.2% | 28.3% | 31.1% |
| MRFalign | **25.5%** | **35.9%** | **39.4%** | **29.7%** | **39.5%** | **43.3%** | **29.4%** | **39.0%** | **43.6%** |

## 3.4 Discussion

In this chapter I have presented a new alignment method that aligns two families through alignment of two Markov Random Fields (MRFs), which models the multiple sequence alignment (MSA) of a protein family using an undirected general graph in a probabilistic way. The MRF representation is better than the extensively-used PSSM and HMM representations in that the former can capture long-range residue interaction pattern, which reflects the overall 3D structure of a protein family. As such, MRF comparison is much more sensitive than HMM comparison in detecting remote homologs. This is validated by our large-scale experimental tests showing that MRF-MRF comparison can greatly improve alignment accuracy and remote homology detection over currently popular sequence-HMM, PSSM-PSSM, and HMM-HMM comparison methods. Our method also has a larger advantage over the others on mainly-beta proteins.

We build our MRF model of a protein family based upon multiple sequence alignment (MSA) in the absence of native structures. The accuracy of the MRF model depends on the accuracy of an MSA. Currently we rely on the MSA generated by PSI-BLAST. In the future, we may explore better alignment methods for MSA building or even utilize solved structures of one or two protein sequences to improve MSA. The accuracy of the MRF model parameter usually increases with respect to the number of non-redundant sequence homologs in the MSA. Along with more and more protein sequences are generated by a variety of sequencing projects, we shall be able to build accurate MRFs for more and more protein families and thus, detect their homologous relationship more accurately.

An accurate scoring function is essential to MRF-MRF comparison. Many different methods can be used to measure node and edge similarity of two MRFs, just like many different scoring functions can be used to measure the similarity of two PSSMs or HMMs. This chapter presents only one of them. In the future we may explore more possibilities. It is computationally intractable to find the best alignment between two MRFs when edge similarity is taken into consideration. In this chapter I also present an ADMM algorithm that can efficiently solve the MRF-MRF alignment problem to suboptimal. However, this algorithm currently is about 10 times slower than the Viterbi algorithm for PSSM-PSSM alignment.

# Chapter 4  A Conditional Neural Fields Model for Protein Alignment

## 4.1 Introduction

In this chapter I will introduce the training and inference algorithm of the node potential introduced in the last chapter. Current protein alignment methods are limited in the following aspects. One is that these methods use linear scoring functions to guide the sequence-template alignment (Eddy, 2001; Söding, 2005). The choice of a scoring function is the key to alignment accuracy. A linear function (Peng and Xu, 2009) cannot deal well with correlation among protein features, although many features are indeed correlated (e.g., secondary structure vs. solvent accessibility). The other issue is that these methods heavily depend on sequence profile. Although sequence profile is very powerful in detecting remote homologs and generating accurate alignments, as demonstrated by HHpred and many others, these method do not work well for many proteins with a very low NEFF which is defined in the previous chapter.

To go beyond the limitations of current alignment methods, in this chapter I will present a novel Conditional Neural Fields (CNF) method for protein alignment used for protein threading, which can align a sequence to a distantly-related template much more accurately. Our method combines homologous information (i.e., sequence profile) and structure information using a probabilistic nonlinear scoring function, which has several advantages over the widely-used linear functions. First, it explicitly accounts for correlations among protein features, reducing over-counting and/or under-counting of protein features. Second, our method can align different regions of the sequence and template using different criteria. For example, we can use sequence information to align disordered regions (since only sequence information is reliable for them) and sequence plus structure information for the others. Third, the relative importance of homologous and structural information is dynamically determined. When proteins under consideration have a sparse sequence profile, we count more on structural information; otherwise homologous information (e.g., sequence profile similarity). Finally, gap probability is estimated using both context specific and position specific features. If protein sequence profile is sparse, we will rely more on context-specific information (i.e., structure information); otherwise the position-specific information derived from alignment of sequence homologs.

The CNF method is able to integrate as much information as possible to estimate the alignment probability of two residues. In particular, the CNF method utilizes neighborhood (sequence and structural) information to estimate the probability of two residues being aligned much more accurately. Neighborhood information is also very helpful in determining gap opening positions. Neighborhood (sequence) information has been used by many programs (e.g., PSIPRED (McGuffin, et al., 2000)) for protein local structure prediction and by few for protein sequence alignment and homology search, but such information has not been applied to protein threading, especially for gap opening. It is much more challenging to make use of neighborhood information in protein threading since it needs to deal with a variety of structure information.

We also use a quality-sensitive method to train the CNF model, as opposed to the standard maximum-likelihood (ML) method. The ML method treats all the aligned positions equally. However, not all the aligned positions in an alignment are equally important. Some positions are more conserved than others and more important for the construction of 3D models from the alignments. By directly maximizing the expected quality of a set of training alignments, the quality-sensitive method weighs more on the conserved positions to ensure accurate alignment. Experimental results confirm that the quality-sensitive method usually can result in better alignments.

Tested on both public (but small) benchmarks and large-scale in-house datasets, our CNF method generates significantly better alignments than the best profile-based method (e.g., HHpred) and several top threading methods including BThreader (Peng and Xu, 2009), Sparks and MUSTER. Our method performs especially well when only distantly-related templates are available or when proteins under consideration have sparse sequence profile.

## 4.2 Methods

**Conditional Neural Fields (CNF) for Protein Threading**

CNF (Conditional Neural Fields) is a recently-developed probabilistic graphical model, integrating and embracing the strength of both Conditional Random Fields (CRFs) (Lafferty, et al., 2001) and neural networks. CNF not only can parameterize conditional probability in the log-linear form (which is similar to CRFs), but also can implicitly model complex/nonlinear relationship between input features and output labels (which is similar to neural networks). CNF has been applied to protein secondary structure prediction (Peng, et al., 2009; Wang, et al., 2011), protein conformation sampling (Zhao, et al., 2010) and handwriting recognition (Peng, et al., 2009). Here we describe how to model protein sequence-template alignment using CNF.

Let $T$ and $S$ denote a template protein with solved structure and a target protein without solved structure, respectively. Each protein is associated with some protein features, e.g., sequence profile, (predicted) secondary structure, (predicted) solvent accessibility. Let $A = \{a_1, a_2, \dots, a_{L_A}\}$ denote an alignment between $T$ and S where $L_A$ is the alignment length and $a_i$ is one of the three possible states $M$, $I_t$ and $I_S$, which we have introduced in the previous chapter. As shown in Figure 9, an alignment can be represented as a sequence of three states and assigned a probability calculated by our CNF model. The alignment with the highest probability is deemed as the optimal. We calculate the probability of one alignment A as follows.

$$P(A|T,S,\theta) = exp\left(\sum_{i=1}^{L_A} E(a_{i-1}, a_i, T, S)\right)/Z(T,S) \tag{13}$$

where θ is the model parameter vector to be trained, i indicates one alignment position and $Z(T,S) = \sum_A exp\left(\sum_{i=1}^{L_A} E(a_{i-1}, a, T, S)\right)$ is the normalization factor (i.e., partition function) summing over all possible alignments for a given protein pair. The function $E$ in Eq. (13) estimates the log-likelihood of state transition from $a_{i-1}$ to $a_i$ based upon protein features. It is a nonlinear scoring function defined as follows.

$$E(a_{i-1}, a_i, T, S) = \varphi(a_{i-1}, a_i, T, S) + \phi(a_i, T, S) \tag{14}$$

Where $\varphi$ and $\phi$ are called edge and node functions introduced in the chapter 3, respectively, quantifying correlation among alignment states and protein features. Both the node and edge feature functions can be as simple as a linear function or as complex as a neural network. Here we use neural networks with only one hidden layer to construct these two types of functions. Due to space limit, we only explain the edge feature function in detail. The node feature function is similar but slightly simpler. Since in total there are 9 possible state transitions in an alignment, we need 9 edge feature functions, each corresponding to one kind of state transition. Figure 9 shows an example of the edge feature function for the state transition from $M$ to $I_t$. Given one state transition $u$ to $v$ at position $i$ where $u$ and $v$ are two alignment states, the edge feature function is defined as follows.

$$\varphi(a_{i-1} = u, a_i = v, T, S) = \sum_j \lambda_{u,v}^j H_{u,v}^j \left(w_{u,v}^j f_{u,v}(T, S, i)\right) \tag{15}$$

Where $f_{u,v}(T, S, i)$ is the feature generation function, which generates input features from the target and template proteins for the alignment at position $i$. The feature generation function is state-dependent, so we may use different features for different state transitions. In Eq. (15), $j$ is the index of the hidden neurons in the hidden layer, $\lambda_{u,v}^j$ is the model parameter between one hidden neuron and the output layer, $H_{u,v}^j(x)$ (sigmoid function) is the gate function for the hidden neuron conducting nonlinear transformation of input, and $w_{u,v}^j$ is the model parameter vector connecting the input layer to one hidden neuron. All the model parameters are state-dependent, but position-independent. In total there are 9 different neural networks for the 9 state transitions. These neural networks have separate model parameters. All of them constitute the model parameter vector $\theta$ introduced in Eq. (13).

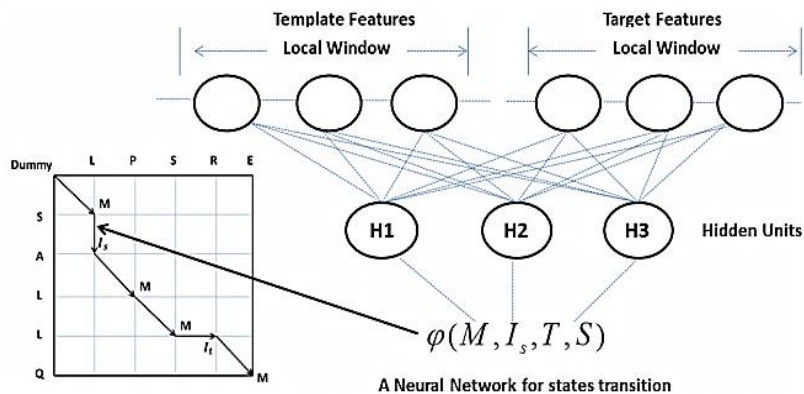

**Figure. 9.** An example of the edge feature function φ, which is a neural network with one hidden layer. The function takes both template and target protein features as input and yields one likelihood score for state transition $M$ to $I_t$. Meanwhile, $H1$, $H2$ and $H3$ are hidden neurons conducting nonlinear transformation of the input features.

Because a hidden layer is introduced to CNF to improve expressive power over CRF, it is important to control the model complexity to avoid over-fitting. We do so by using a *L2-norm* regularization factor to restrict the search space of model parameters. This regularization factor is determined by 5-fold cross validation. Once the CNF model is trained, we can calculate the optimal alignment using the Viterbi algorithm (Forney Jr, 1973).

**Training CNF Model by Quality-Sensitive Method**

CRFs/CNFs are usually trained by maximum likelihood (ML) or maximum a posteriori (MAP). The ML method trains the CRFs/CNFs model parameters by maximizing the occurring probability of a set of reference alignments, which are built by a structure alignment tool. The ML method treats all the aligned positions equally, ignoring the fact that some are more conserved than others. It is important not to misalign the conserved residues since they may be related to protein function. As such, it makes more sense to treat conserved and non-conserved residues separately. Although there are a few measures for the degree of conservation to be studied, here we simply use the local TM-score (Xu and Zhang, 2010) between two aligned residues. Given a reference alignment (and the superimposition of two proteins in the alignment), the local TM-score at one alignment position i is defined as follows.

$$w_i = \frac{1}{1+(d_i/d_0)^2} \tag{16}$$

Where $d_i$ is the distance deviation between the two aligned residues at position $i$ and $d_0$ is a normalization constant depending on only protein length. TM-score ranges from 0 to 1 and the higher the more conserved the aligned position is. When the alignment state at position $i$ is gap, the local TM-score is equal to 0 and $w_i$ is equal to 0 at a gap position.

To differentiate the degree of conservation in the alignment, we train the CNF model by maximizing the expected TM-score. The expected TM-score of one threading alignment is defined as follows.

$$Q = \frac{1}{N(A)} \sum_i (w_i MAG_i) \tag{17}$$

Where N(A) is the smaller length of the two proteins and $MAG_i$ is the marginal alignment probability at alignment position i. Since $w_i$ is equal to *0* at a gap position, Eq. (17) de facto sums the marginal alignment probabilities at all the alignment positions with the match state (i.e., state *M*). Given two residues of a protein pair, their marginal alignment probability is equal to the accumulative probability of all the possible alignments of this protein pair in which these two residues are aligned to each other. The marginal alignment probability can be calculated efficiently using the forward-backward algorithm (Lafferty, et al., 2001). Eq. (17) is similar to the definition of TM-score except that the latter does not have a term for the marginal alignment probability. By maximizing Eq. (17), we weigh more on those aligned residue pairs with higher local TM-score (i.e., more conserved residue pairs) instead of treating all the aligned residue pairs equally. We term this method as quality-sensitive training method. The central problem of quality-sensitive training method is to calculate the gradient of Eq. (17). Next I will introduce the algorithm details.

Given an alignment, $A = \{a_1, a_2, ..., a_{L_A}\}$, let $A[i, L_A] = \{a_i, a_{i+1}, ..., a_{L_A}\}$ denote a left-partial alignment starting with the *N*-terminal end to position $i$ and let $A[i, L_A] = \{a_i, a_{i+1}, ..., a_{L_A}\}$ denote a right-partial alignment starting from the *C*-terminal end to position $i$. Let $x$ and $y$ denote the number of target and template residues contained in the left-partial alignment $A[1, i]$, respectively. Both $x$ and $y$ can also be treated as the residue indices in the target and template proteins, respectively. Therefore, each alignment position index $i$ is associated with a pair of residue indices $x$ and $y$. Let $m$ and $n$ denote the number of residues in the target and template proteins, respectively. In total there are $mn$ possible pairs of residue indices. Note that when alignment position $i$ corresponds to a pair of residue indices $x$ and $y$, the alignment position $i - 1$ may correspond to one of the three possible residue index pairs $(x-1, y-1)$, $(x, y-1)$ or $(x-1, y)$, depending on the alignment state at position $i$.

**Gradient calculation.** Let $F_i^v$ denote the accumulative probability of all possible left-partial alignments ending at alignment position $i$ with state $v$. Similarly, let $B_i^u$ denote the accumulative probability of all possible right-partial alignments ending at alignment position $i$ with state $u$. $F_i^v$ and $B_i^u$ are the forward and backward functions, which have been extensively described in the sequence alignment literature. Sometimes, we also write $F_i^v$ as $F_{x,y}^v$ or $B_i^u$ as $B_{x,y}^u$ when it is necessary to explicitly spell out the residue indices. Both $F_i^v$ and $B_i^u$ can be calculated recursively as follows.

$$F_i^v = \sum_u F_{i-1}^u \exp(E(a_{i-1}=u, a_i=v, S, T)) \qquad (18)$$

$$B_i^u = \sum_v B_{i+1}^v \exp(E(a_i=u, a_{i+1}=v, S, T)) \qquad (19)$$

The marginal alignment probability $MAG_i$ can be calculated as follows.

$$MAG_i = \frac{F_i^M B_i^M}{Z} \qquad (20)$$

Meanwhile, the normalization factor $Z$ (i.e., partition function) is equal to $\sum_u F_i^u B_i^u$ for any $i$. In particular, we have

$$Z = \sum_u F_{m,n}^u = \sum_u B_{1,1}^u \qquad (21)$$

$MAG_i$ depends on the model parameter $\theta$, so we only need to calculate $\frac{\partial MAG_i}{\partial \theta}$ in order to calculate the gradient. Based upon Eq. (20), we have

$$\frac{\partial MAG_i}{\partial \theta} = \frac{\partial}{\partial \theta}\left(\frac{F_i^M B_i^M}{Z}\right) = \frac{\partial F_i^M}{\partial \theta}\frac{B_i^M}{Z} + \frac{\partial B_i^M}{\partial \theta}\frac{F_i^M}{Z} - \frac{F_i^M B_i^M}{Z}\frac{\partial Z}{\partial \theta} \qquad (22)$$

Since $Z = \sum_u F_{m,n}^u$, we have $\frac{\partial Z}{\partial \theta} = \sum_u \frac{\partial F_{(m,n)}^u}{\partial \theta}$. That is, $\frac{\partial MAG_i}{\partial \theta}$ depends on only $\frac{\partial F_i^M}{\partial \theta}$ and $\frac{\partial B_i^M}{\partial \theta}$. For the purpose of simplicity, let $E_i^{u \to v}$ denote $E(a_{i-1}=u, a_i=v, S, T)$ (see Eq. (13-15)). By Eq. (18), we have

$$\frac{\partial F_i^v}{\partial \theta} = \sum_u \left(\frac{\partial F_{i-1}^u}{\partial \theta}\exp(E_i^{u \to v}) + \exp(E_i^{u \to v})\frac{\partial E_i^{u \to v}}{\partial \theta}F_{i-1}^u\right) \qquad (23)$$

Eq. (23) indicates that $\frac{\partial F_i^v}{\partial \theta}$ can be calculated recursively. Similarly, $\frac{\partial B_i^u}{\partial \theta}$ can also be calculated recursively. Since $E_i^{u \to v}$ is a neural network, $\frac{\partial E_i^{u \to v}}{\partial \theta}$ can be calculated using the gradient chain rule with time complexity depending on the architecture of the neural network. The size of the neural network is determined by the number of features, the window size, and the number of hidden neurons, but independent of protein length. As such, we can assume the size is a large constant. There are, in total, $mn$ possible residue index pairs for the alignment position $i$ in $F_i^v$ and $B_i^u$, so the time complexity of the gradient calculation is $O(mn)$, the product of the target and template protein lengths.

The expected TM-score in Eq. (17) is not concave, so it is challenging to optimize it to globally optimal. Here we use the L-BFGS (Limited memory BFGS (Malouf, 2002)) algorithm to solve it to suboptimal. To obtain a good solution, we run L-BFGS several times starting from different initial solutions and use the best suboptimal solution as the final. In order to use the L-BFGS algorithm, we need to calculate the gradient of Eq. (17), which is detailed in the supplementary file. In addition, we obtained the best performance when using 12 hidden neurons in the hidden layer for all the different neural networks in both edge and label functions in Eq. (14).

**Protein Features**

We generate position-specific score matrix (PSSM) for a template and position specific frequency matrix (PSFM) for a target using PSI-BLAST with five iterations and E-value 0.001. Let $PSSM(i, aa)$ denote the mutation potential for amino acid aa at template position $i$ and $PSFM(j, aa)$ the occurring frequency of amino acid aa at target position $j$.

**Features for match state.** We use the following features to estimate the alignment probability of two residues:

**Sequence profile similarity**. The profile similarity between two positions is calculated as $\sum_a PSSM(i,a)PSFM(j,a)$. In addition, we also calculate sequence similarity using the Gonnet matrix (Gonnet, et al., 1992) and BLOSUM62.

**Amino acids substitution matrix**. We use two matrices. One is the matrix developed by Kihara group (Tan, et al., 2006) and the other is a structure-based substitution matrix. Each entry in Kihara's matrix measures the similarity between two amino acids using the correlation coefficient of their contact potential vectors. The contact potential vector of one amino acid contains 20 elements, each indicating the contact potential with one of the 20 amino acids. The structure-based substitution matrix (Tan, et al., 2006) is more sensitive than BLOSUM for the alignment of distantly-related proteins.

**Secondary structure score**. The secondary structure similarity between the target and template is evaluated in terms of both the 3-class and 8-class types. We generate secondary structure types of the template using DSSP (Kabsch and Sander, 1983). We also predict the 3-class and 8-class secondary structure types for the target using PSIPRED and our in-house tool RaptorX-SS8 (Wang, et al., 2011), respectively.

**Solvent accessibility score**. We discretize the solvent accessibility into three equal-frequency states: buried, intermediate and exposed. The equal-frequency method is the best among several discretization methods we tested. We use our in-house tool to predict the solvent accessibility state of the target and DSSP to calculate the template solvent accessibility. Let sa denote the solvent accessibility type on the template. The solvent accessibility similarity is defined as the predicted likelihood of the target residue being in sa.

**Environment fitness score.** This score measures how well to align one sequence residue to a specific template environment. The environment of a template residue is defined as the combination of its solvent accessibility state and 3-class secondary structure type, which results in 9 environment types.

**Neighborhood similarity score.** It was shown that conserved positions tend to cluster together along the sequence. That is, if two residues can be aligned it is likely that the residues around them can also be aligned. Therefore, we can use the neighborhood information to estimate the likelihood of two residues being aligned. The neighborhood information used in our model includes sequence profile, secondary structure and solvent accessibility in a window of size 11.

**Residues in two terminals.** Residues at the two terminals may not have sufficient neighborhood information and thus, needs some special handling. We use one binary variable to indicate if one residue is at the two terminals or not.

**Disordered region.** Disordered regions are natively unfolded or intrinsically unstructured, lacking stable tertiary structure and the predicted accuracy at those regions is very low. Therefore, we cannot use the structure information (e.g., secondary structure and solvent accessibility) to align/un-align disordered regions because it might introduce more false positive. We use DISOPRED (Ward, et al., 2004) to predict disordered regions in a sequence, which also produces a confidence score, ranging from 0 to 9, to indicate how likely a residue is in a disordered region. The higher confidence score, the more likely a residue being in a disordered region. We deem a residue to be in a disordered region if the confidence score is 9. When aligning disordered residues, only sequence information is used and all the predicted structure information is ignored (i.e., their relevant feature values are set to 0).

**Features for gap state.** We do not use an affine gap penalty, which is being extensively used in many sequence alignment programs. Instead, we use both position-specific and context-specific features to estimate gap probability. The position-specific features are derived from the sequence homologs of a given protein while the context-specific features include amino acid identify, hydropath index, both 3-class and 8-class secondary structure and solvent accessibility. We also use a head and tail gap feature to indicate if a gap appears at the two terminals or in the middle of a protein.

## 4.3 Results

**Training data.** We constructed the training and validation data sets from PDB25 downloaded from PISCES (Wang and Dunbrack, 2003). Any two proteins in PDB25 shares less than 25% sequence identity. A set of 1010 protein pairs is built from PDB25 as the training data, which covers most of the SCOP fold classes. Another set of 200 protein pairs is constructed from PDB25 as our validation data. There is no redundancy between the training and validation data sets (i.e., <25% sequence identity). The structure alignments (i.e., reference alignments) for the training and validation data are built by our in-house structure alignment tool DeepAlign.

**Test data.** We use the following 4 test sets.

1. In-House benchmark. A large in-house set consists of 3600 protein pairs from PDB25. This set has no redundancy with our training and validation data (i.e., < 25% sequence identity). It is constructed so that 1) it contains all protein classes (alpha, beta and alpha-beta proteins); 2) the protein NEFF values are almost uniformly distributed; 3) the protein length is widely distributed; and 4) TM-scores of all the pairs are spread out between 0.5 and 0.7.

2. MUSTER benchmark (Wu and Zhang, 2008). It contains all the training data used by the MUSTER threading program, consisting of 110 hard ProSup pairs and another 190 pairs selected by Zhang group, each pair having TM-score > 0.5.

3. SALIGN benchmark (Braberg, et al., 2012), It contains 200 protein pairs, each of which shares ~20% sequence identity and ~65% of structurally equivalent residues with RMSD <3.5 Å. Many protein pairs in this set contain proteins of very different size, which makes it very challenging for any threading methods.

4. ProSup benchmark (Lackner, et al., 2000). This set consists of 127 protein pairs. Programs to compare. We compare our CNF threading method, denoted as CNFpred, with the top-notch profile-based and threading methods such as HHpred, SPARKS/SP3/SP5, SALIGN, RAPTOR and BThreader. We use the published results for SPARKS/SP3/SP5 since they have their own template file formats and we cannot correctly run the programs locally. We use the published result for SALIGN since it is unavailable. We also compare our program with MUSTER. We focus on comparing our CNFpred with HHpred and BThreader since the latter two performed extremely well in the most recent CASP competition in 2010.

**Evaluation criteria.** We evaluate the threading methods using both reference-dependent and reference-independent alignment accuracy. The reference-dependent accuracy is defined as the percentage of correctly aligned positions judged by the reference alignments. For all the benchmarks we use 4 structure alignment tools to generate reference alignments, including TM-align, Matt, Dali (Holm and Sander, 1995) and our in-house structure alignment tool DeepAlign. Besides, the ProSup and MUSTER benchmarks have their own reference alignments. To avoid bias towards a specific structure alignment tool, we evaluate threading alignment accuracy using all of the reference alignments mentioned above. Note that our CNFpred is trained using only the structure alignments generated by our in-house tool DeepAlign. To evaluate the reference-independent alignment accuracy, we build a 3D model for the target protein using MODELLER (Eswar, et al., 2006; Fiser and Šali, 2003) from its alignment to the template and then evaluate the quality of the resultant 3D model using TM-score. TM-score ranges from 0 to 1, indicating the worst and best model quality, respectively.

**Table 13.** Reference-dependent alignment accuracy on the In-House benchmark. Columns 2-5 indicate four different reference alignment generation tools. Bold indicates the best performance.

| Methods | TMalign | Dali | Matt | DeepAlign |
|---|---|---|---|---|
| HHpred(Local) | 32.63 | 36.60 | 35.53 | 35.47 |
| HHpred(Global) | 38.80 | 43.65 | 42.48 | 42.78 |
| BThreader | 37.44 | 41.85 | 40.17 | 40.95 |
| CNFpred | 37.44 | 51.77 | 49.98 | 51.19 |

**Reference-dependent alignment accuracy.** As shown in Tables 13-17, CNFpred outperforms all the others regardless of the reference alignments used. The advantage of CNFpred over the popular profile-profile alignment method HHpred increases with respect the hardness of the benchmark. For example, on the most challenging In-House benchmark the relative improvement of CNFpred over HHpred is more than 20%. Even on the easiest ProSup benchmark the relative improvement of CNFpred over HHpred is ~10%. Our old threading program BThreader works well on the ProSup and SALIGN sets, but not as well on the MUSTER and In-House benchmarks. On these two benchmarks, BThreader has similar performance as HHpred (global alignment), but much worse than CNFpred. CNFpred has a smaller advantage on SALIGN because it contains many proteins with symmetric domains, which have several good alternative alignments. However, we only use the first alignments generated by the structure alignment tools as the reference alignments. Therefore, even if the threading alignments are pretty good, they may still have very low accuracy when judged by the "non-perfect" reference alignments.

**Table 14.** Reference-dependent alignment accuracy on the MUSTER benchmark. Columns 2-5 indicate four different reference alignment generation tools. Column "BR" indicates the reference alignments provided in the benchmark. The result of the program MUSTER is the training accuracy taken from (Wu and Zhang, 2008). All the other numbers are test accuracy. Bold indicates the best performance.

| Methods | TMalign | Dali | Matt | DeepAlign | BR |
|---|---|---|---|---|---|
| HHpred(Local) | 42.96 | 57.34 | 46.00 | 46.50 | 45.34 |
| HHpred(Global) | 48.82 | 53.13 | 51.48 | 52.48 | 51.48 |
| BThreader | 47.35 | 51.30 | 50.13 | 50.53 | 50.01 |
| CNFpred | 54.17 | 58.46 | 57.26 | 59.14 | 57.06 |

**Table 15.** Reference-dependent alignment accuracy on the SALIGN benchmark. The results of SALIGN, RAPTOR and SPARKS, SP3 and SP5 are taken from (Peng and Xu, 2009; Xu, et al., 2003). Bold indicates the best performance. Columns 2-5 correspond to four different reference alignment generation tools.

| Methods | TMalign | Dali | Matt | DeepAlign |
|---|---|---|---|---|
| SPARKS | 53.10 | - | - | - |
| SALIGN | 56.40 | - | - | - |
| RAPTOR | 40.00 | - | - | - |
| SP3 | 56.30 | - | - | - |
| SP5 | 59.70 | - | - | - |
| HHpred(Local) | 60.64 | 62.94 | 62.97 | 63.16 |
| HHpred(Global) | 62.98 | 63.14 | 63.87 | 63.53 |
| BThreader | 64.40 | 63.13 | 63.05 | 64.09 |
| CNFpred | 66.73 | 67.95 | 68.17 | 69.50 |

**Table 16.** Reference-dependent alignment accuracy on the ProSup benchmark. Columns 2-5 correspond to four different reference alignment generation tools. Column "BR" denotes the reference alignment provided in the benchmark. The results of SALIGN, RAPTOR and SPARKS, SP3 and SP5 are taken from (Peng and Xu, 2009; Xu, et al., 2003). Bold indicates the best performance.

| Methods | TMalign | Dali | Matt | DeepAlign | BR |
|---|---|---|---|---|---|
| SPARKS | - | - | - | - | 57.2 |
| SALIGN | - | - | - | - | 58.3 |
| RAPTOR | - | - | - | - | 61.3 |
| SP3 | - | - | - | - | 65.3 |
| SP5 | - | - | - | - | 68.7 |
| HHpred(Local) | 57.53 | 60.58 | 60.61 | 60.36 | 64.90 |
| HHpred(Global) | 61.84 | 65.31 | 64.52 | 65.29 | 69.04 |
| BThreader | 60.87 | 64.89 | 63.97 | 64.26 | 76.08 |
| CNFpred | **66.26** | **71.16** | **71.06** | **72.01** | **77.09** |

**Table 17.** Reference-independent alignment accuracy, measured by TM-score, on the four benchmarks: In-House, MUSTER, SALIGN and ProSup. The result of the program MUSTER is its training accuracy (Wu and Zhang, 2008). All the other numbers are test accuracy. Bold indicates the best performance.

| Methods | In-House | MUSTER | SALIGN | ProSup |
|---|---|---|---|---|
| HHpred (Local) | 1047.56 | 108.84 | 119.97 | 53.88 |
| HHpred (Global) | 1522.77 | 142.00 | 121.83 | 56.44 |
| MUSTER | - | 136.47 | - | - |
| BThreader | 1537.89 | 143.95 | 132.85 | 66.77 |
| CNFpred | **1692.17** | **152.14** | **134.50** | **67.34** |

**Reference-independent alignment accuracy.** Tested on the much more challenging In-House benchmark, we obtain a TM-score of 1693, which is more than 10% better than HHpred and BThreader In addition, on SALIGN and Prosup, our CNFpred obtains total TM-score of 134.5 and 67.34, respectively. By contrast, HHpred has TM-score of 121.83 and 56.44, respectively, as shown in Table 17.

**Alignment accuracy with respect to the sparsity of sequence profile.** To further examine the performance of CNFpred, BThreader and HHpred with respect to the amount of homologous information, we divide the protein pairs in the In-House benchmark into 9 groups according to the minimum NEFF value of a protein pair and calculate the average TM-score of the target models in each group. As shown in Figure 10 (A), when NEFF is small (i.e., proteins have sparse sequence profile), our method exceeds BThreader and HHpred significantly. We also divide all the protein pairs in the benchmark into 4 groups according to the NEFF values of two proteins in a pair using the threshold 6. As shown in Fig. 10 (B), when the NEFF values of both proteins in a pair are small (<6), our method is 33% better than HHpred. When one of the two proteins in a pair has NEFF less than 6, our method is 25% better than HHpred. Even when both proteins in a pair have NEFF values larger than 6, which indicates both proteins have sufficient homologous information, our method still outperforms HHpred slightly. In summary, our method is especially good for proteins with sparse sequence profile.

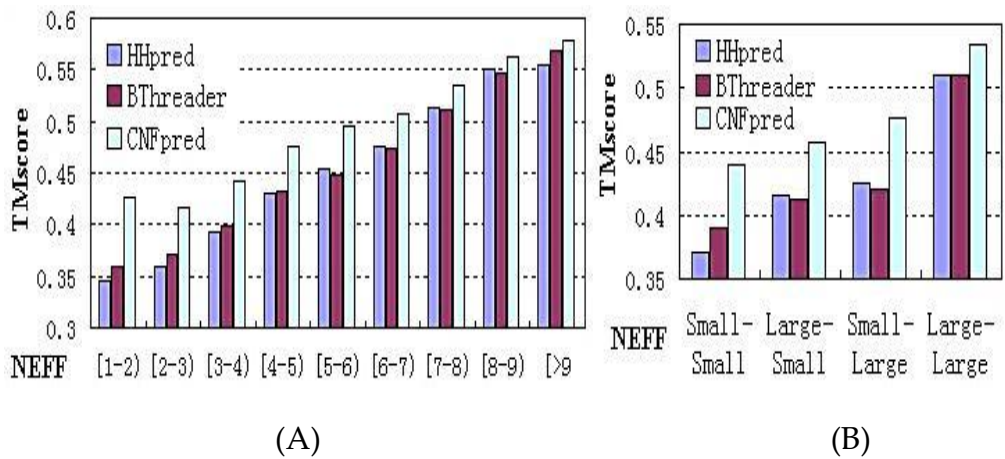

(A)  (B)

**Figure. 10.** Reference-independent alignment accuracy with respect to sparsity of sequence profile (i.e., NEFF). (A) NEFF is divided into 9 bins. (B) NEFF is divided into two bins at the threshold 6.

**Alignment accuracy with respect to protein classes and lengths.** As shown in Figure 11, our method is superior to others across all protein classes and length. Our method does especially well for all-beta proteins because our method can make use of structure information in a better way.

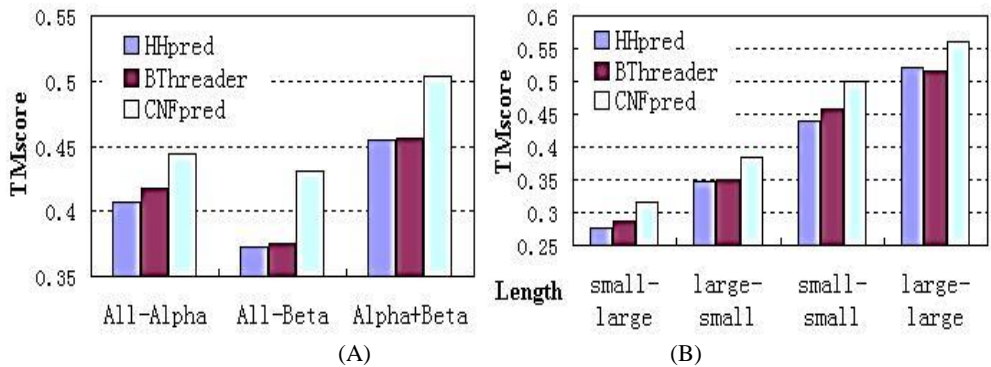

(A)  (B)

**Figure. 11.** (A) Reference-independent alignment accuracy with respect to (A) protein class and (B) protein length. A protein with less than 150 amino acids is treated as small; otherwise as large.

**Threading performance with different features.** Table 17 lists the alignment quality measured by TM-score when different protein features are used. The results are obtained by training with the same data mentioned before and testing on our In-House benchmark. This benchmark is challenging so that it is easier to show the difference of different features and training approaches. For Maximum

Likelihood and quality-sensitive training, we run Viterbi algorithm to generate alignments and for Maximum A Posteriori we run Maximum Expected Accuracy (MAC) algorithm (Söding, 2005) to generate alignments. We evaluate the contribution of 8-class second structure and 3-class solvent accessibility and treat sequence profile and 3-class secondary structure as the base control. As is shown in Table 18, 8-class secondary structure and 3-class solvent accessibility improves the alignment accuracy by 0.01 and 0.02, respectively, no matter which training approaches are used. The predicted solvent accessibility improves the alignment accuracy the most. The MAP training method improves the accuracy by 0.01 over the Maximum Likelihood method while the quality-sensitive method improves the accuracy by 0.01 over the MAP method. The quality-sensitive method significantly improves the overall performance.

**Table 18.** The contribution of adding different features and using different training methods testing on our In-House benchmark.

| Feature | Maximum Likelihood | Maximum A Posteriori | Quality-Sensitive |
|---|---|---|---|
| Profile + SS3[a] | 1536.01 | 1578.60 | 1612.44 |
| +SS8[b] | 1567.81 | 1595.72 | 1637.64 |
| +SA[c] | 1606.68 | 1616.04 | 1662.12 |
| +SS8 + SA[d] | 1633.32 | 1664.28 | 1692.17 |

a Using profile relating and 3-class secondary structure features
b Using profile relating and 3-class secondary structure features plus 8-class secondary structure features.
c Using profile relating and 3-class secondary structure features plus solvent accessibility features.
d Using profile relating and 3-class secondary structure features plus both 8-class secondary structure and solvent accessibility features.

**Threading performance on a large set.** We tested CNFpred and HHpred on a fairly large set constructed from PDB25. All the ~6000 proteins in PDB25 are used as templates and 1000 of them are randomly chosen as the target set. We run CNFpred and HHpred to predict the 3D structure for each of the 1000 targets using all the ~6000 templates. We run HHpred using his "realign" option. By using that HHpred first run local alignment to search the database and then use global alignment to re-align templates and targets. When predicting structure for one specific target protein, the target itself is removed from the template list. After generating alignments for a specific target, CNFpred ranks the templates using a neural network, which predicts the quality (i.e., TM-score) of an alignment. The template with the best predicted quality is used to build a 3D model for the target. HHpred is run with the default options. As shown in Fig. 12 (A), CNFpred is significantly better than HHpred when the targets are not so easy (i.e., the HHpred model has TM-score <0.7). On the 1000 targets, CNFpred and HHpred obtain overall TM-score of 558 and 515, respectively. If we exclude the 170 easy targets (i.e., either CNFpred or HHpred model has TM-score >0.8) from consideration, the overall TM-score obtained by CNFpred and HHpred are 416 and 375, respectively. That is, CNFpred is ~10.9% better than HHpred. As shown in Fig. 12(B), the number of targets for which CNFpred generates models better than HHpred by at least 0.05 TM-score is 329 while HHpred is better than CNFpred by this margin for only 76 targets. Furthermore, the number of targets for which CNFpred generates models better than HHpred by at least 0.10 TM-score is 192 targets while HHpred is better than CNFpred by this margin for only 27 targets. In summary, CNFpred has very large advantage over HHpred on hard targets.

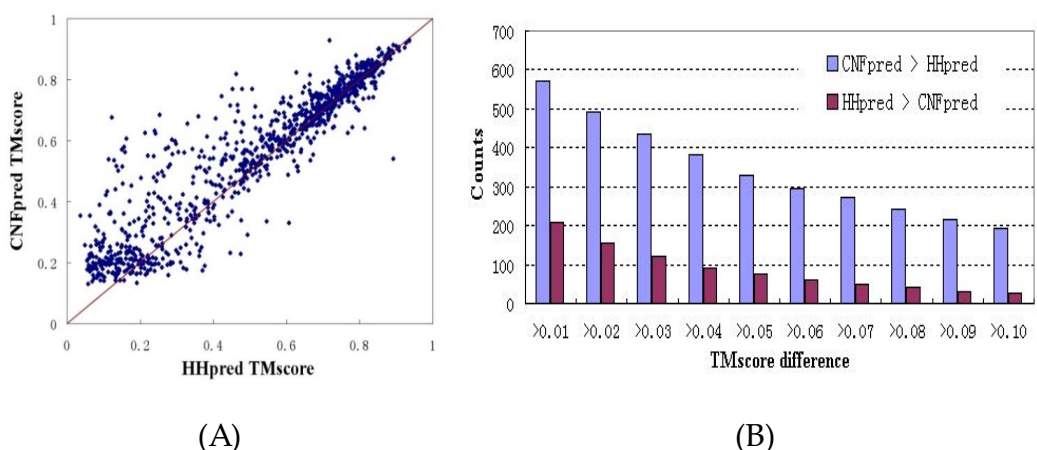

(A)            (B)

**Figure. 12.** (A) TM-scores of the CNFpred and HHpred models for the 1000 targets from PDB25. Each point represents 2 models generated by CNFpred and HHpred, respectively. (B) Distribution of the TM-score difference of two 3D models for the same target. Each blue (red) column shows the number of targets for which CNFpred (HHpred) is better by a given margin.

## 4.4 Conclusion

In this chapter I has presented a novel conditional neural fields (CNF) model for protein threading of proteins with sparse sequence profile. Our CNF method can take advantage of as many correlated sequence and structure features as possible to improve alignment accuracy. We also presents a quality-sensitive training method to improve alignment accuracy, as opposed to the standard maximum-likelihood method. Although using many features and a nonlinear scoring function, our CNF method still can efficiently generate the optimal alignments by dynamic programming. It takes only seconds to thread a typical protein pair. Experimental results demonstrate that our CNF method can greatly improve alignment accuracy over other CASP-winning programs, regardless of benchmarks, reference alignments, protein classes and lengths. Currently, our CNF model only considers state transition between two adjacent positions. We can also model pairwise interaction between two non-adjacent positions, but it is computationally challenging to train such a model. Some approximation algorithms may be resorted.

Homologous information is very effective in detecting remote homologs, as evidenced by the profile-based method HHpred, which performed better than many threading methods in recent CASP events. It shows that homologous information is not sufficient for proteins with sparse sequence profile (i.e., low NEFF). Our method can improve alignment accuracy over profile-based methods by using more structure information, especially for proteins with sparse sequence profile. Cowen group takes a rather different method, called simulated evolution, to enrich sequence profile and shows that the alignment accuracy can be improved for some proteins (Kumar and Cowen, 2009). The capability of predicting structures for proteins with sparse sequence profile is very important. Simple statistics indicate that among the ~6877 Pfam families (Bateman, et al., 2004) without solved structures, 79.2%, 63.7%, 45.5% and 25.4% have NEFF ≤ 6, 5, 4, and 3, respectively, and of the 5332 Pfam families with solved structures, ~57% have NEFF <6. In addition, ~25% of the protein sequences in Uniprot (Consortium, 2011) are not covered by Pfam (ver 25.0). A significant number of these sequences are singletons (i.e., products of orphan genes) and, thus, have NEFF=1. Many low-homology proteins and families (NEFF ≤ 6) will continue to be lack of solved structures in the foreseeable future. Therefore, our CNF threading method shall be useful for a large percentage of protein sequences without solved structures.

# Chapter 5   Protein Alignments Using Context-Specific Potential

## 5.1  Introduction

In this chapter I will introduce a novel context-specific alignment potential for protein threading including both alignment and template selection. Our alignment potential measures the log odds ratio of one alignment being generated from two related proteins to being generated from two unrelated proteins, by integrating context-specific information. An alignment is assumed to be optimal if it maximizes its potential. The alignment potential quantifies how well one sequence residue can be aligned to one template residue based upon context-specific information of these two residues. We need a potential function because we need to eliminate the bias caused by background probability. As shown in Figure 13, proper alignment potential function can eliminate the bias introduced by the training datasets as well as sequence length. Generally speaking, the alignment potential function tends to keep the relatively conserve aligned regions and makes shorter alignments. This makes the ranking algorithm select the templates much easier.  Experimental results show that our context-specific alignment potential is much more sensitive than the widely used context-independent or profile-based (which is position-specific) scoring function, generating significantly better alignments and threading results than the best profile-based methods on several very large benchmarks. Our method works particularly well for distantly-related proteins or proteins with sparse sequence profiles due to the effective integration of context-specific, structure and global information.

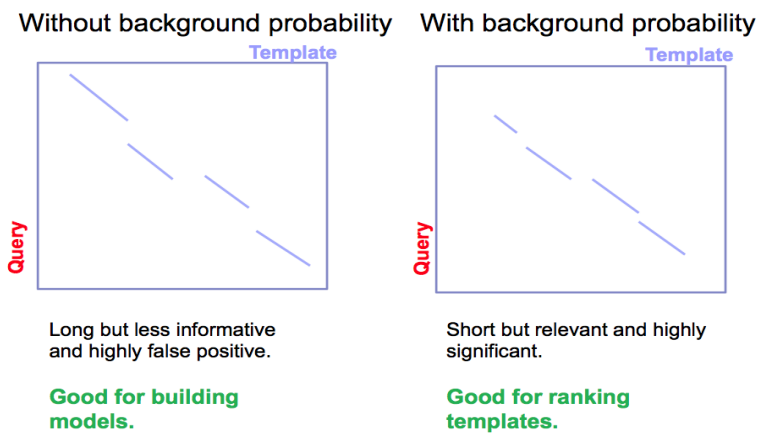

**Figure 13.** Different protein alignments under different alignment scoring functions. The left one does not consider the background probability. The right one uses an alignment potential function considering the background probability.

## 5.2 Methods

**Protein Alignment Potential Function**

We represent one alignment A between two proteins as a sequence of alignment states $a_1, a_2, \ldots, a_L$ where L is the alignment length and $a_i$ is the alignment state at position $i$. There are three possible alignment states $M$, $I_t$ and $I_s$ where $M$ represents two residues being aligned, $I_t$ denotes an insertion in the template, and $I_s$ denotes an insertion in the sequence.

Similar to many amino acid substitution matrices such as BLOSUM and PAM, which defines the mutation potential of two amino acids, we define the potential of one protein alignment. Given a protein sequence $S$ and a template $T$ and one of their alignments $A$, let $P(A|S,T)$ denote the probability of A being generated from $S$ and $T$ using our alignment method. We define the potential of $A$, denoted as $U(A|S,T)$, as follows.

$$U(A|S,T) = log \frac{P(A|S,T)}{P_{ref}(A)} \qquad (24)$$

Where $P_{ref}(A)$ is the background (or reference) probability of $A$, i.e., the probability of $A$ being generated from two randomly-selected proteins with the same lengths as $S$ and $T$, respectively. Intuitively, an alignment is good as long as its probability is much better than the expected probability. We assume that an alignment is optimal if it maximizes its potential. That is, given a sequence and a template, we can find their optimal alignment by maximizing the alignment potential function.

Here $P(A|S,T)$ can be any function calculating the probability of alignment $A$ given two proteins $S$ and $T$. For example, it can be the Conditional Neural Field model introduced in Chapter 4. It can also be the probabilistic graphical model of MRFalign containing the long distance correlation between alignment states. $P(A|S,T)$ can be defined as follows.

$$P(A|T,S,\theta) = \frac{exp(F(A|T,S,\theta)+w \cdot G(A|T,S,\theta))}{Z(T,S,\theta)} \quad (25)$$

Where $\theta$ is the model parameter vector to be trained, $w$ (=1.0) is a weight factor and $Z(T,S,\theta)$ is the normalization factor (i.e., partition function) summing over all possible alignments for a given protein pair. For the purpose of simplicity, we omit $\theta$ in the following sections unless we have to spell it out. The function $F$ estimates the log-likelihood of one sequence residue being aligned to one template residue based upon the input protein features. The function G estimates the log-likelihood of a pair of sequence residues being placed into two template positions at a given distance based upon the input protein features. The functions $F$ and $G$ are just the node alignment potential function and edge alignment function introduced in Chapter 3.

We can train their parameters by maximum-likelihood. That is, given a set of training protein pairs and their reference alignments (built by a structure alignment tool), we maximize their occurring probability defined by Eq. (25). However, since $G(A|T,S)$ is a global alignment function, it is computationally hard to directly maximize Eq. (25). In addition, it may cause over-fitting by training the parameters of $F$ and $G$ simultaneously since the parameter space is very big. To avoid these problems we determine the parameters of functions $F$ and $G$ separately, which will be explained in the following sections.

**Reference state.** We can calculate the reference alignment probability $P_{ref}(A)$ in Eq. (24) by randomly sampling a set of protein pairs, each with the same lengths as the sequence $S$ and template $T$, respectively, and then estimating the probability of alignment $A$ based upon these randomly sampled protein pairs. As long as we generate sufficient number of samplings, we shall be able to approximate $P_{ref}(A)$ very well. Here we use the geometric mean to approximate the reference state as follows.

$$P_{ref}(A) = \sqrt[N]{\prod_{i=1}^{N} P(A|X,Y)} \quad (26)$$

Where N is the number of samplings and $X$ and $Y$ represent two sampled proteins with the same lengths as $S$ and $T$, respectively. Combining Eq. (24-26), the potential of one alignment $A$ can be calculated as follows.

$$U(A|S,T) = \log \frac{P(A|S,T)}{P_{ref}(A)}$$
$$= \log \frac{P(A|S,T)}{\sqrt[1/N]{\prod_{i=1}^{N} P(A|X,Y)}}$$
$$= \log \frac{exp(F(A|T,S)+w\cdot G(A|T,S))/Z(S,T)}{\sqrt[N]{\prod_{i=1}^{N} exp(F(A|X,Y)+w\cdot G(A|X,Y))/Z(X,Y)}} \qquad (27)$$

Note that an alignment is represented as a sequence of 3 states (match $M$, insertion at sequence $I_s$ and insertion at template $I_t$). Therefore, given two sequence-template pairs $(S,T)$ and $(X,Y)$, as long as $S$ and $T$ have the same lengths as $X$ and $Y$, respectively, the alignment space (i.e., the set of all possible alignments) for $S$ and $T$ is same as that for $X$ and $Y$. That is, any $S$-to-$T$ alignment is also a feasible alignment between $X$ and $Y$, although it may have different probabilities. Conversely, any $X$-to-$Y$ alignment is also a feasible alignment between $S$ and $T$.

By definition, $Z(S,T)$ is equal to the alignment space size times the mean value of the denominator in Eq. (25). Since $S$ and $T$ have the same alignment space as $X$ and $Y$, $Z(S,T)$ differs from $Z(X,Y)$ only in the mean values of their corresponding denominators in Eq. (25), which is independent of any specific alignment, but may depend on protein residue composition.

Therefore, we have

$$U(A|S,T) = \left(F(A|T,S) - EXP_{X,Y}F(A|X,Y)\right)$$
$$+w \cdot \left(G(A|T,S) - EXP_{X,Y}G(A|X,Y)\right) + c(S,T) \qquad (28)$$

Where EXP is the expectation operator, $F(A|T,S) - EXP_{X,Y}F(A|X,Y)$ can be interpreted as node alignment potential and $G(A|T,S) - EXP_{X,Y}G(A|X,Y)$ as edge alignment potential.

In Eq. (28), $c(S,T)$ depends on only the residue composition of S and T but not any specific alignment. In particular, $c(S,T)$ is equal to 0 if the sampled protein pairs have similar residue composition as $S$ and $T$. As such, for the purpose of finding the optimal alignment between $S$ and $T$, we can simply ignore $c(S,T)$. Therefore, the key challenge is to determine the node and edge alignment potential functions in the right hand side of Eq. (25).

## 5.3 Results

**Training and validation data.** We constructed the training and validation data from BC40, a subset of PDB, in which any two proteins shares less than 40% sequence identity. In total we use a set of 1800 protein pairs as the training data, which covers most of the folds in the SCOP database, and a set of 500 protein pairs as the validation data. There is no redundancy between the training and validation data (i.e., <40% sequence identity). The training and validation data has the following properties: 1) All the proteins have length less than 400 and contain less than 10% of residues with missing coordinates; 2) The TM-score of a protein pair is uniformly distributed from 0.55 to 1; and 3) We use our in-house structure alignment tool DeepAlign to generate the reference alignment for a protein pair. Each alignment has fewer than 50 middle gaps and the number of terminal gaps is <20% of the alignment length.

**Test data for alignment.** We use the following 3 datasets to test the alignment accuracy of our method.

1. Set6K: a set of ~6000 protein pairs. Any two target proteins in this set share <40% sequence identity. The TM-score of a protein pair is uniformly distributed between 0.55 and 0.8. Two proteins in a pair have small length difference. The protein pairs in Set6K have 5% of overlap with our training and validation data. By "overlap" we mean that the proteins in one pair have sequence identity 30-50% with those in another pair.

2. Set4K: a set of 4547 protein pairs. Any two target proteins in the set share < 25% sequence identity. The protein pairs in Set4K have 3% of overlap with our training and validation data. Two proteins in a pair have length difference larger than 30%, so this set can be used to test if the domain boundary is correctly aligned or not.

3. Set180K: a very large set of 179,390 protein pairs. Any two proteins in most pairs share <40% sequence identity. The TM-score of a protein pair is uniformly distributed between 0 and 1. Note that the size of our training set is only 1% of this large set, so the test result on this set is unlikely biased by the training set.

Test data for threading. We use the following 2 datasets to test the threading accuracy of our method.

Set1000×6000: a large set constructed from PDB25, which consists of ~6000 proteins. All the proteins in PDB25 are used as templates and 1000 of them are randomly chosen as the target proteins. We predict the 3D structure for all the 1000 targets using the ~6000 templates, but excluding self-threading.

CASP10: a set of 123 test proteins. We use the CASP official domain boundary definition for each test protein.

**Evaluation criteria and programs to compare.** We evaluate our threading method using both reference-dependent and reference-independent alignment accuracy. The reference-dependent accuracy is defined as the percentage of correctly aligned positions judged by the reference alignments, which are built using our in-house tool DeepAlign. We also built the reference alignments using other structure alignment tools such as DALI, Matt and TMalign and observed similar performance trend. To evaluate the reference-independent alignment accuracy, we build a 3D model for the target protein using MODELLER from its alignment to the template and then evaluate the quality of the resultant 3D model using TM-score. TM-score ranges from 0 to 1, indicating the worst and best model quality, respectively. Since our ultimate goal is to predict 3D structure for a target protein, reference-independent alignment accuracy is more important than reference-dependent accuracy. We compare our method with the top-notch profile-based method HHalign, which is run with the option "-mact 0.1".

**Table 19.** Reference-dependent (Ref-dep) and reference-independent (Ref-ind) alignment accuracy on two benchmarks Set6K and Set4K. Reference-independent alignment accuracy is measured by TM-score.

|  | Set6K | | Set4K | |
| --- | --- | --- | --- | --- |
|  | Ref-dep | Ref-ind$^{TM}$ | Ref-dep | Ref-ind$^{TM}$ |
| Our work | 52% | 0.52 | 63% | 0.62 |
| HHalign | 45% | 0.44 | 57% | 0.56 |

**Table 20.** Reference-dependent alignment accuracy on two benchmarks of Set6K and Set4K.

|  | Set6K | | Set4K | |
| --- | --- | --- | --- | --- |
|  | Ref-dep | 4-position off | Ref-dep | 4-position off |
| Our work | 52% | 57% | 63% | 67% |
| HHalign | 45% | 50% | 57% | 60% |

As shown in Table 19, our method outperforms HHalign in terms of both reference-dependent and reference-independent alignment accuracy on the two benchmarks Set6K and Set4K. On these two sets, our method outperforms HHalign by 13.6% and 9%, respectively, in terms of the model quality (i.e., reference-independent accuracy). However, in terms of reference-dependent accuracy, our method is better than HHalign by only 8.8% and 5.2%, which is not as big as reference-independent accuracy. To find out why, we calculate the reference-dependent accuracy on Set6K and Set4K by allowing 4-position off the exact match, as shown in Table 20, which indicates that our method is much better than HHalign in terms of the reference-dependent accuracy when 4-position off the exact match is allowed. This may explain why our method can yield much better 3D model quality.

**Table 21.** Reference-dependent (Ref-dep) and reference-independent (Ref-ind) alignment accuracy on the very large benchmark Set180K. Reference-independent alignment accuracy is measured by TM-score. All the protein pairs are divided into 4 groups depending on their structure similarity measured by TM-score.

|          | Ref-dependent | | Ref-independent$^{TM}$ | |
| --- | --- | --- | --- | --- |
| TM-score | HHalign | Our work | HHalign | Our work |
| 0.80-1.00 | 83% | 84% | 0.78 | 0.79 |
| 0.65-0.80 | 60% | 62% | 0.52 | 0.56 |
| 0.40-0.65 | 32% | 35% | 0.30 | 0.34 |
| 0.25-0.40 | 11% | 19% | 0.16 | 0.20 |

As shown in Table 21, on the very large Set180K set, our method yields slightly better performance than HHalign when two proteins under consideration are very similar. This is not surprising since most methods can generate pretty good alignments for two closely-related proteins. When the TM-score of two proteins under consideration falls into (0.65, 0.80), our method outperforms HHalign by ~3.3% in terms of the reference-dependent accuracy and by ~7.6% in terms of the reference-independent accuracy. When the TM-score of two proteins under consideration falls into (0.40, 0.65), our method outperforms HHalign by ~9.4% in terms of the reference-dependent accuracy and by ~11.3% in terms of the reference-independent accuracy.

When the TM-score of two proteins under alignment falls into (0.25, 0.40), our method outperforms HHalign by a very large margin in terms of reference-dependent alignment accuracy. However, in terms of the reference-independent alignment accuracy, the advantage of our method is not as big, although it is still substantial. This may be because that MODELLER cannot build a reasonable model from an alignment with too many errors. That is, when the TM-score of two proteins is less than 0.4, it may not be so important to generate an accurate alignment for them since the resultant 3D model has low quality and thus, will not be very useful.

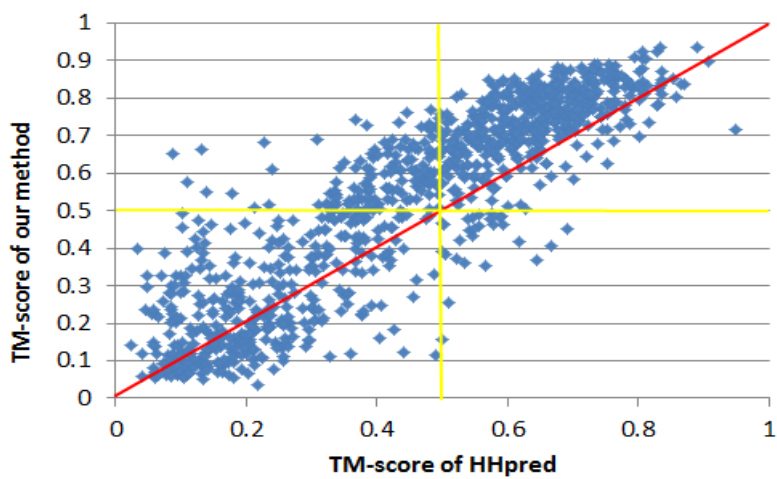

**Figure. 14.** The quality of the models by our method and HHpred for the 1000 targets randomly chosen from PDB25. Each point represents two models generated by our method (Y-axis) and HHpred (X-axis), respectively.

Threading performance on a large test set. We test the threading performance of our method and HHpred on Set1000×6000. We run both our method and HHpred to predict the 3D structure for each of the 1000 targets using the ~6000 templates. HHpred is run with its "realign" option. That is, HHpred first searches through the template database using local alignment and then re-aligns a target to the top templates using global alignment. By doing so, HHpred can improve its modeling accuracy a little bit over the default mode. To speed up, our method first aligns a target to all the templates using only the local alignment potential and then ranks all the templates using both the local and global alignment potentials described in section Methods. After ranking only the first-ranked templates are used to build a 3D model by MODELLER for each target.

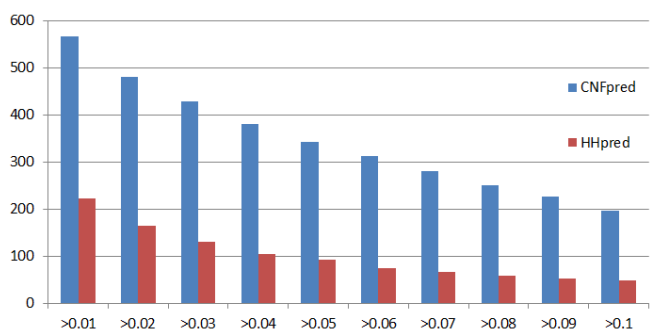

**Figure. 15.** Distribution of the model quality difference, measured by TM-score. Each blue column shows the number of targets for which our method is better by a given margin. Each red column shows the number of targets for which HHpred is better by a given margin.

As shown in Figure 14, our method is significantly better than HHpred when the targets are not so easy (i.e., the HHpred model has TM-score<0.7). On the 1000 targets, our method and HHpred obtain average TM-score 0.566 and 0.517, respectively. Our method outperforms HHpred no matter whether the target is easy or hard. If we exclude the 170 easy targets (i.e., either our model or HHpred model has TM-score>0.8) from consideration, the accumulative TM-score obtained by our method and HHpred are 0.524 and 0.451, respectively. That is, our method is ~16.1% better than HHpred. Further, as indicated by the yellow lines in Figure 14, our method can generate models with TM-score >0.5 for many targets for which HHpred fails to generate a model with TM-score >0.5. We use TM-score=0.5 as a cutoff because when a model has TM-score>0.5, its overall fold is basically correct.

As shown in Figure 15, our method generates models better than HHpred by at least 0.05 for 342 targets while HHpred is better than our method by this margin for only 93 targets. Further, the number of targets for which our method generates models better than HHpred by at least 0.10 is 197 while HHpred is better than our method by this margin for only 49 targets. In summary, our method has a large advantage over HHpred on hard targets.

Threading performance on CASP10 data set. We further evaluate the threading performance of our method on the most recent CASP10 targets. We use the CASP official domain boundary definition for each target and in total there are 123 test proteins. To make the test as fair as possible, both our method and HHpred used the same set of templates and the same protein sequence database (i.e., NR), which were constructed before CASP10 started.

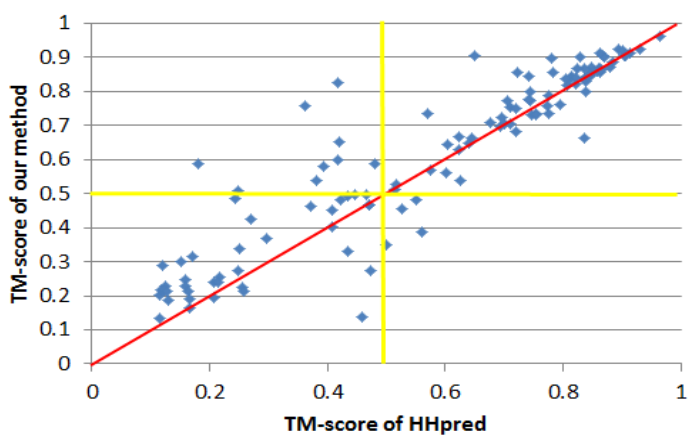

**Figure. 16.** The model quality, measured by TM-score, of our method and HHpred for the 123 CASP10 targets. Each point represents two models generated by our method (y-axis) and HHpred (x-axis), respectively.

As shown in Figure 16, similar to what we have observed on the large threading test set, our method significantly outperforms HHpred when the targets are not so easy. Our method generates a model with TM-score >0.5 for quite a few targets for which HHpred fails to generate a model with TM-score>0.5. On the whole test set, our method and HHpred obtain accumulative TM-score 77.52 and 70.65, respectively. If we exclude the "Server" targets from consideration and only look at the more challenging "Human/Server" targets. The average TM-score obtained by our method and HHpred are 0.63 and 0.57, respectively. That is, our method is ~10.5% better than HHpred.

It is very challenging to fairly compare our single-template threading method with the CASP10-participarting servers because that most of the CASP10 servers used a hybrid method instead of an individual threading method. For example, the first-ranked Zhang-Server integrated both consensus analysis of ~10 individual threading programs and fragment-based 3D model building technique. The top-ranked HHpred server integrated new profile generation method, multi-template alignment and a better 3D model building technique. The top-ranked Robetta server used consensus results from three programs including HHpred, RaptorX and SPARKS and also a very new 3D model building method (citation here later). Our RaptorX server, which is ranked No. 2 overall, employed multiple-template threading, which can generate better 3D models than single-template threading for many targets especially the easy ones. In summary, the accumulative TM-score obtained by our single-template threading method described in this paper is only 0.85 less than what was obtained by RaptorX in CASP10. It can be ranked No.6 among all the CASP10-participating servers.

**P-value.** It is desirable that any structure prediction program can assign a confidence score to predicted models. Here we use P-value to quantify the relative quality of the top-ranked templates and alignments. To calculate the P-value, we employs a set of reference templates, which consists of ~1800 single-domain templates with different SCOP folds. Given a target, we first thread it to this reference template database and then estimate an extreme value distribution from the ~1800 alignment scores (i.e., alignment potentials). Based upon this distribution, we calculate the P-value of each alignment when threading the target to the real template database. The P-value actually measures the quality of the template (and the alignment) by comparing it to the reference templates.

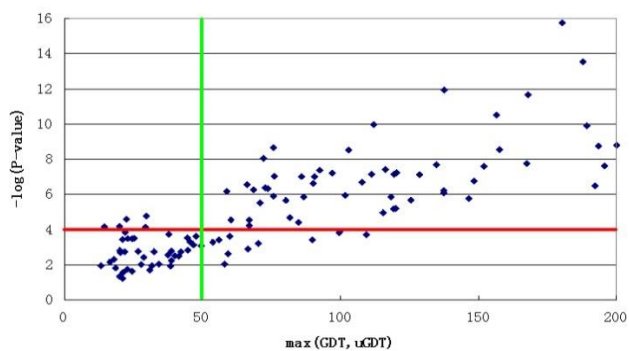

**Figure. 17.** The relationship between P-value and the model quality on the 123 CASP10 targets. The x-axis is the model quality measured by max(GDT, uGDT) and the y-axis is –log(P-value).

To measure the real model quality, we use both GDT (Global Distance Test) (Zemla, et al., 1999) and uGDT (i.e., un-normalized GDT). GDT has been employed as an official measure by CASP for many years. It measures the quality of a model by comparing it with the native and outputs a value from 0 to 100, indicating the worst and the best quality, respectively. uGDT is equal to GDT times the target length divided by 100. uGDT is more suitable when the target protein is relatively large or multi-domain and there are only good templates covering a segment of the target (e.g., one of the domains). In this case, GDT is likely to be quite small and not a good indicator even if the good templates are identified since GDT is normalized by the whole target length. However, uGDT is not good for a target with length<100. For example, when a target of 60 residues is covered by a template perfectly on 48 of the 60 residues, the uGDT of this alignment is 48 while the GDT is 80. In this case, GDT is more suitable than uGDT. In summary, we use max(uGDT, GDT) to measure the model quality. We say one alignment is reasonable when its resultant model has uGDT or GDT larger than 50. We use 50 as a cutoff because that many proteins similar at only the fold level have GDT or uGDT around 50.

As shown in Figure 17, the P-value is a reliable indicator of model quality. When P-value is small (i.e. <10-5), the models have uGDT or GDT greater than or equal to 50. Even if P-value is less than 10-4, there are few models have both uGDT and GDT less than 50. That is, the prediction from our threading method is reliable when the P-value is less than 10-4.

**Table 22.** Contribution of pairwise potential to alignment accuracy, tested on two benchmarks Set6K and Set4K. Reference-independent alignment accuracy is measured by TM-score

|  | Set6K | | Set4K | |
| --- | --- | --- | --- | --- |
|  | Ref-dep | Ref-ind$^{TM}$ | Ref-dep | Ref-ind$^{TM}$ |
| Local potential | 49% | 0.51 | 60% | 0.61 |
| Local + Pairwise | 52% | 0.52 | 63% | 0.62 |

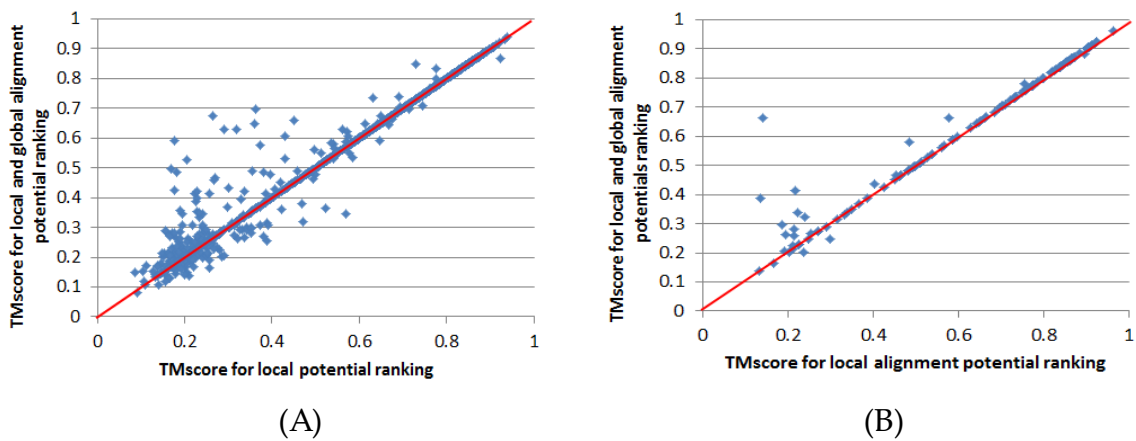

| (A) | (B) |

**Figure. 18.** (A) Contribution of the distance-based pairwise alignment potential to Set1000×6000. Each point represents the quality, measured by TM-score, of 2 models: one is generated using the local alignment potential only (*X*-axis) and the other using both the local and global alignment potentials (*Y*-axis). (B) Contribution of the distance-based pairwise alignment potential to the CASP10 set. Each point represents the quality, measured by TM-score, of 2 models: one is generated using the local alignment potential only (*X*-axis) and the other using both the local and global alignment potentials (*Y*-axis).

**Contribution of the distance-based pairwise potential.** To evaluate the contribution of our pairwise potential to alignment accuracy, we calculate the accuracy improvement resulting from using both our local and pairwise potential over using only our local potential on two benchmarks Set6K and Set4K. As shown in Table 22, our pairwise potential can improve reference-dependent accuracy by 3% and reference-independent accuracy by 0.01, respectively.

We also evaluate the contribution of our pairwise potential to template selection. To speed up, we generate alignments using our local alignment potential and then rank all the templates using a linear combination of our local and pairwise alignment potentials (with equal weight). Experimental results on the 1000×6000 threading set and the CASP10 set indicate that the pairwise potential indeed improves the threading performance, as shown in Figures 18 (A) and (B). On the 1000×6000 set, the average TM-score increases from 0.547 to 0.566 when the pairwise potential is used to rank the templates. On the CASP10 set, the accumulative TM-score increases from 75.58 to 77.52 when the pairwise potential is used. As shown in Figures 18 (A) and (B), the context-specific pairwise potential is particularly helpful to hard targets.

**Case Study.** Here we use two cases to further demonstrate the strength of our method. Both of these two cases are from our Set6K benchmark. The first case is to align two proteins 3qnrA and 2gffA, which have TM-score between 0.62-0.65 according to the structural alignments generated by TMalign, Matt, Dali and our in-house tool DeepAlign. That is, these two proteins are similar in structures but not very much in sequences. Meanwhile, 2gffA contains two alpha and two beta segments which are very similar to one of the two domains of 3qnrA. As shown in Figure 19, our method can correctly align >90% of the positions judged by the reference alignment (regardless of which structural alignment tools are used to generate it). In contrast, HHalign fails to align the 2nd alpha and beta segments. This is partially because HHalign favors generating short alignment. If we choose 3qnrA as the template to build a 3D model for 2gffA, the resultant models from our method and HHalign have TM-score 0.63 and 0.25, respectively.

```
Our Method                                                              HHalign
>3qnrA                                                                  >3qnrA
--GSHMPGPVARLAPQAVLTPPSAASLFLVLVAGDSDDDRATVCDVISGIDGPLKAVGFRELAGSLSCVVGVGAQFWDRVS   GSHMPGPVARLAPQAVLTPPSA-ASLFLVLVAGDSDDDRATVCDVISGIDGPLKAVGFRELAGSLSCVVGV---------
ASSKPAHLHPFVPLSGPVHSAPS-TPGDLLFHIKAARKDLCFEL-----GRQIVSALGSAATVVDEVHGFRY--------                                      -GAQFWDRVSASSKPAHLHPFVPLSGPVH
----FDSRDLLGFVDGTENPTDDDAADSALIGDEDPDFRGGSYVIV                          SAPSTPGDLLFHIKAARKDLCFELGRQIVSALGSAATVVDEVHGFRYFDSRDLLGFVDGTENPTDDDAADSALIGDEDPDFR
                                                                        GGSYVIV
>2gffA                                                                  >2gffA
MH----------------------VTLVEINVKED--KVDQFIEVFRANHLGSIREA-----GNLRFDVLRDE-------                   -------------------MHVTLVEINVKEDK--VDQFIEVFRANHLGSIR-----EAGNLRFDVLRDEHIPTRFYI
----------------------HIPTRFYIYEAYTDEAAVAIHKTTPHYLQCVEQLAPLMTGPRKKTVFIGLMPGSLEHH   YEAYTDEAAVAIHKTTPHYLQCVEQLAPLMTGPRKKTVFIGLMPGSLEHHHHHH--------------------------
HHHH--------------------------------                                    -------
```

**Figure. 19.** Two alignments between 3qnrA and 2gffA generated by our method and HHalign. The blue and red colors demonstrate correctly-aligned regions judged by the reference alignment. To save space, only one of the domains of 3qnrA is shown.

Here we use another two proteins 3k53A and 1cb7A to show the case that our method and HHalign generate two alignments of nearly the same length but our alignment has much better quality. As shown in Figure 20, our method aligns nearly 80% of positions correctly while HHalign fails to align any position correctly. If we use 3k53A as the template to build models for 1cb7A, the resultant 3D models from our method and HHalign have TM-score 0.64 and 0.22, respectively. We can also examine the alignments visually. As shown in Fig. 21 (A) and (B), our method aligns the local structure very well while HHalign seemingly produces a totally wrong alignment. In this case both 3k53A and 1cb7A have pretty good sequence profile information and the predicted secondary structure for 1cb7A is also very accurate (>80%).

```
Our method                                                          HHalign
>3k53A                                                              >3k53A
--MVLKTVALVGNPN---VGKTTIFNALTGLRQHVGNWPGVTVEKKEGIMEYRE-KEFLVVDLPGIYSLTAHSIDEL IARNF   MVLKTVALVGNPNVGKTTIFNALTGLRQHVGNWPGVTVEKKEGIMEYREKEFLVVDLPGIYSLTAHSIDELIARNFILDGN-
ILDGNADVIVDIVDST-CLMRNLFLTLELFEMEVKNIILVLNKFDLLKKKGAKIDIKKMRKELGVPVIPTNAKKGEGVEELK   --ADVIVDIVDSTCLMRNLFLTLELFEMEVKNIILVLNKFDLLKKKGAKIDIKKMRKELGVPVIPTNAKKGEGVEELKRMIA
RMIALMAE---GKVTNPIIPRYDEDIEREIKHISELLRGTPLAEKYPIRWLALKLLQRDEEVIKLVLKYLGQEKMDEILKHI   LMAEGKVTNPI--------------------------------------IPRYDEDIEREIKHISELLR
SELEEKYKRPLDIVIASQKYEFLEQLLRKFVVHE                                  GTPLAEKYPIRWLALKLLQRDEEVIKLVLKYLGQEKMDEILKHISELEEKYKRPLDIVIASQKYEFLEQLLRKFVVHE

>1cb7A                                                              >1cb7A
ME--KKTIVLGVIGSDCHAVGNKILDHAFT-----------------------NAGFNVVNIGVLSP-------QELFIKA   -----------------------------------------------------------------
AIETKADAILVSSLYGQGEIDCKGLRQKCDEAGLEGILLYVGGNIVVG-KQHWPDVEKRFKDMGYDRVYAP----GTPPEVGI   MEKKTIVLGVIGSDCHAVGNKILDHAFTNAGFNVVNIG--VLS----PQ-ELFIKAAIETKADAILVSSLYGQGEIDCKGL
ADLKKDLNIE---------------------------                               RQKCDEAGLEGILLYVGGNIVVGKQHWPDVEKRFKDMGYDRVYAPGTPPEVGIADLKKDLNIE-----------
                                                                    ---------------------
```

**Figure. 20.** The alignments between 3k53A and 1cb7A generated by our method and HHalign. The blue and red colors indicate the correctly aligned regions.

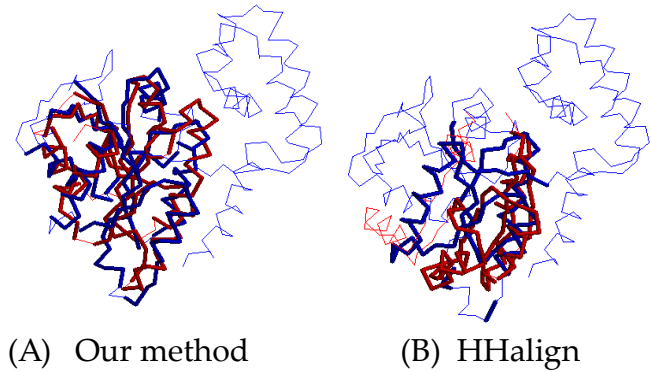

(A) Our method  (B) HHalign

**Figure. 21.** Superposition between protein 3k53A and 1cb7A based on the alignments generated by our method and HHalign, respectively. The highlighted regions indicate the aligned residues.

## 5.4 Conclusion

In this chapter I presented a novel context-specific alignment potential which measures the log odds ratio of one alignment being generated from two related proteins to being generated from two unrelated proteins. It can be used for both protein alignment and protein threading, Intuitively, an alignment is regarded as good only when its estimated probability is much higher than the expected. Our alignment potential uses context-specific and structure information through advanced machine learning techniques such as Conditional Neural Fields, which can combine a variety of highly-correlated protein sequence and structure features, without worrying too much about over-counting and under-counting of features. Experimental results show that our context-specific alignment potential is much more sensitive than the widely-used context-independent (e.g. profile-based) scoring function and yields significantly better alignments and threading results. Our method works particularly well for distantly-related proteins or proteins with sparse sequence profiles due to the effective integration of context-specific, structure and global information.

This chapter also shows that our context-specific distance-based pairwise potential is helpful to protein threading, as opposed to the contact-based potentials previously used by some protein threading methods. Combined with our context-specific local alignment potential, our distance-based pairwise potential can help improve both alignment accuracy and template selection especially for hard targets.

# Chapter 6    Protein Contact Prediction by Integrating Joint Evolutionary Coupling Analysis and Supervised Learning

## 6.1  Introduction

Protein contacts contain important information for protein folding and recent works indicate that one correct long-range contact for every 12 residues in the protein allows accurate topology level modeling (Kim, et al., 2014). Thanks to high-throughput sequencing and better statistical and optimization techniques, evolutionary coupling (EC) analysis for contact prediction has made good progress, which makes de novo prediction of some large proteins possible.

Nevertheless, contact prediction accuracy is still low even if only the top *L*/10 (L is the sequence length) predicted contacts are evaluated.

Existing contact prediction methods can be roughly divided into two categories: 1) evolutionary coupling (EC) analysis methods, such as, that make use of multiple sequence alignment; and 2) supervised machine learning methods, such as SVMSEQ (Wu and Zhang, 2008), NNcon (Tegge, et al., 2009), SVMcon (Cheng and Baldi, 2007), CMAPpro (Di Lena, et al., 2012), that predict contacts from a variety of information including mutual information, sequence profile and some predicted structure information. In addition, a couple of methods also use physical constraints, such as PhyCMAP (Wang and Xu, 2013) and Astro-Fold (Klepeis and Floudas, 2003).

Residue EC analysis is a pure sequence-based method that predicts contacts by detecting co-evolved residues from the MSA (multiple sequence alignment) of a single protein family. This is based upon an observation that a pair of co-evolved residues is often found to be spatially close in the 3D structure. Mutual information (MI) is a local statistical method used to measure residue co-evolution strength, but it cannot tell apart direct and indirect residue interaction and thus, has low prediction accuracy. Along with many more sequences are available, some global statistical methods, such as maximum entropy and probabilistic graphical models, are developed to infer residue co-evolution from MSA. These global statistical methods can differentiate direct from indirect residue couplings and thus, are more accurate than MI. See (de Juan, et al., 2013) for an excellent review of EC analysis. Representative tools of EC analysis include Evfold (Marks, et al., 2011), PSICOV (Jones, et al., 2012), GREMLIN (Kamisetty, et al., 2013), and plmDCA (Ekeberg, et al., 2013). Meanwhile, GREMLIN and plmDCA do not assume Gaussian distribution of a protein family. All these EC methods make use of residue co-evolution information only in the target protein family, ignoring other related families.

Supervised machine learning methods make use of not only mutual information (MI), but also sequence profile and other protein features, as opposed to EC analysis that makes use of only residue co-evolution. Experiments show that due to use of more information, supervised learning may outperform EC methods especially for proteins with few sequence homologs. Recently, a few groups such as DNcon (Eickholt and Cheng, 2012), CMAPpro (Di Lena, et al., 2012) and PConsC (Skwark, et al., 2013) have also applied deep learning, an emerging supervised learning method, to contact prediction and showed some improved performance.

In this chapter, I will present a new method CoinDCA (Ma, et al., 2015) (co-estimation of inverse matrices for direct-coupling analysis) for contact prediction that integrates joint multi-family EC analysis and supervised machine learning. Since joint EC analysis and supervised learning use different types of information, their combination shall lead to better prediction accuracy. The contribution of this paper lies in two aspects. First, different from existing EC analysis that makes use of residue co-evolution information only in the target protein family, our joint EC analysis predicts contacts of a target family not only using its own residue co-evolution information, but also those in its related families which may share similar contact maps. By enforcing contact map consistency in joint EC analysis, we can greatly improve contact prediction. To fulfill this, we develop a statistical method called group graphical lasso (GGL) to estimate the joint probability distribution of a set of related families and enforce contact map consistency proportional to evolutionary distance. Second, we use Random Forests, a popular supervised learning method, to predict the probability of two residues forming a contact using a variety of evolutionary and non-evolutionary information. Then we integrate the predicted probability as prior into our GGL formulation to further improve the accuracy of joint EC analysis. In Chapter 3 I mentioned that MI and DI could be used as features in our framework to improve the quality of protein alignments. CoinDCA can also be treated as a new kind of information source to advance protein alignments.

Experiments show that our method greatly outperforms existing EC or supervised machine learning methods regardless of the number of sequence homologs available for a target protein under prediction, and that our method not only performs better on conserved contacts, but also on family-specific contacts. We also find out that contact prediction may be worsened by merging multiple related families into a single one followed by single-family EC analysis, or by consensus of single-family EC analysis results.

## 6.2 Methods

**Probabilistic Model of a Single Protein Family**

Modeling a single protein family using a probabilistic graphical model has been described in a few papers (Cheng and Baldi, 2007; Jones, et al., 2012; Marks, et al., 2011). Here we briefly introduce it since it is needed to understand our joint graphical model. Given a protein family k and the MSA (multiple sequence alignment) of its sequences, let X denote this MSA where $X_{ir}^k$ is a 21-dimension binary vector indicating the amino acid type (or gap) at row $r$ (of this MSA) and column $i$ and $X_{ir}^k(a)$ is equal to 1 if the amino acid at row $r$ (of this MSA) and column $i$ is a. Let $\bar{X}_i^k$ denote the mean vector of $X_{ir}^k$ across all the rows (i.e., proteins). Let $L$ denote the sequence length of this family and $N_k$ the number of sequences. Assuming this MSA has a Gaussian distribution $N(\mu^k, \Sigma^k)$ where $\mu^k$ is the mean vector with $21L$ elements and $\Sigma^k$ the covariance matrix of size $21L \times 21L$. The covariance matrix consists of $L^2$ sub-matrices, each having size $21 \times 21$ and corresponding to two columns in the MSA. Let $\Sigma_{ij}^k$ denote the sub-matrix for columns $i$ and $j$. For any two amino acids (or gap) $a$ and $b$, their corresponding entry $\Sigma_{ij}^k(a,b)$ can be calculated as follows.

$$\Sigma_{ij}^k(a,b) = \frac{1}{N_k}\sum_{r=1}^{N_k}(X_{ir}^k(a) - \bar{X}_i^k(a))(X_{jr}^k(b) - \bar{X}_j^k(b)) \tag{29}$$

The $\Sigma^k$ calculated by Eq. (29) actually is an empirical covariance matrix, which can be treated as an estimation of the true covariance matrix. Let $\Omega^k = (\Sigma^k)^{-1}$ denote the inverse covariance matrix (also called precision matrix), which indicates the residue or column interaction (or co-evolution) pattern in this protein family. In particular, the zero pattern in $\Omega^k$ represents the conditional independence of the MSA columns. Similar to $\Sigma_{ij}^k$, the precision sub-matrix $\Omega_{ij}^k$ indicates the interaction strength (or inter-dependency) between two columns $i$ and $j$, which are totally independent (given all the other columns) if only if $\Omega_{ij}^k$ is zero.

Due to matrix singularity, we cannot directly calculate $\Omega^k$ as the inverse of the empirical covariance matrix. Instead, we may estimate $\Omega^k$ by maximum-likelihood with a regularization factor $\lambda_1$ as follows.

$$\max_{\Omega^k} logP(X^k|\Omega^k) - \lambda_1\|\Omega^k\|_1$$

Where $\|\Omega^k\|_1$ is the $L1$ norm of $\Omega^k$, which is used to make $\Omega^k$ sparse. Since $P$ is Gaussian, the above optimization problem is equivalent to the following.

$$\max_{\Omega^k}\left(log|\Omega^k| - tr(\Omega^k\hat{\Sigma}^k)\right) - \lambda_1\|\Omega^k\|_1$$

Where $\hat{\Sigma}^k$ is the empirical covariance matrix calculated from the MSA. The PSICOV method for contact prediction is based upon the above formulation.

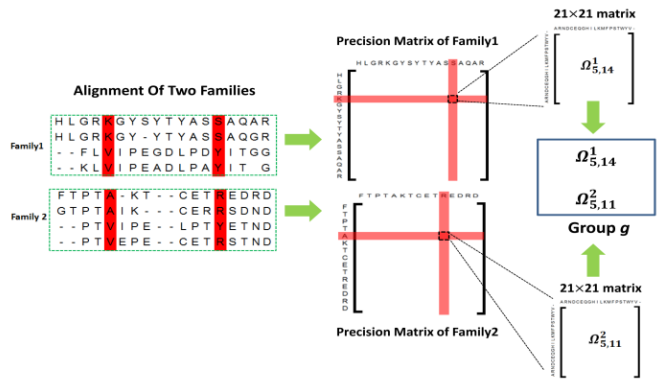

**Figure 22.** Illustration of column pair and precision sub-matrix grouping. Columns 5 and 14 in the 1st family are aligned to columns 5 and 11 in the 2nd family, respectively, so column pair (5,14) in the 1st family and the pair (5,11) in the 2nd family are assigned to the same group. Accordingly, the two precision sub-matrices $\Omega^1_{5,14}$ and $\Omega^2_{5,11}$ belong to the same group.

**Probabilistic model of multiple related protein families by Group Graphical Lasso (GGL)**

The previous section introduces how to model a single protein family using a Gaussian graphical model (GGM). In this section I present our probabilistic model for a set of K related protein families using a set of correlated GGMs. Here we still assume that each protein family has a Gaussian distribution with a precision matrix $\Omega^k$ ($k = 1, 2, \ldots, K$). Let $\Omega$ denote the set $\{\Omega^1, \Omega^2, \ldots, \Omega^K\}$ and $X = \{X^1, X^2, \ldots, X^K\}$ denote the set of MSAs. If we assume that the K families are independent of each other, we can estimate their precision matrices by maximizing their joint log-likelihood as follows.

$$\max_{\Omega} \ \log P(X|\Omega) = \log \prod_{k=1}^{K} P(X^k|\Omega^k) - \lambda_1 \sum_{k=1}^{K} \|\Omega^k\|_1$$

$$= \sum_{k=1}^{K} \left( \log|\Omega^k| - tr(\Omega^k \hat{\Sigma}^k) \right) - \lambda_1 \sum_{k=1}^{K} \|\Omega^k\|_1 \qquad (30)$$

To model the correlation of these families, we assume that the precision matrices are correlated. Now we will show how to model the correlation of the precision matrices through the alignment of these protein families.

We build a multiple sequence alignment (MSA) of these K protein families using a sequence alignment method. Each column in this MSA may consist of columns from several families. If column pair $(j_1, j_3)$ in family $k_1$ is aligned to column pair $(j_2, j_4)$, the interaction strength between two columns $j_1$ and $j_3$ in family $k_1$ shall be similar to that between columns $j_2$ and $j_4$ in family $k_2$. That is, if there is one contact between two columns $j_1$ and $j_3$, then it is very likely there is also a contact between two columns $j_2$ and $j_4$. Accordingly, the precision sub-matrix $\Omega_{j1,j3}^{k1}$ for the two columns $j_1$ and $j_3$ in the family $k_1$ shall be related to the sub-matrix for the two columns $j_2$ and $j_4$ in the family $k_2$ (i.e., $\Omega_{j2,j4}^{k2}$). The correlation strength between $\Omega_{j1,j3}^{k1}$ and $\Omega_{j2,j4}^{k2}$ depends on the conservation level of these two column pairs. That is, if these two column pairs are highly conserved, $\Omega_{j1,j3}^{k1}$ and $\Omega_{j2,j4}^{k2}$ shall also be highly correlated. Otherwise, they may be only weakly related. Based upon this observation, we divide all the column pairs into groups so that any two aligned column pairs belong to the same group, as shown in Figure 22. Therefore, if a target family has L columns aligned to at least one auxiliary family, then there are in total $L(L-1)/2$ groups.

Let G denote the number of groups and K the number of involved families. We estimate the K precision matrices by taking into account their correlation using group graphical lasso (GGL) as follows.

$$max \sum_{k=1}^{K} \left( log|\Omega^k| - tr(\Omega^k \hat{\Sigma}^k) \right) - \lambda_1 \sum_{k=1}^{K} ||\Omega^k||_1 - \sum_{g=1}^{G} \lambda_g ||\Omega_g||_2 \qquad (31)$$

Where g represents one group and $||\Omega_g||_2 = \sqrt{\sum_{(i,j,k)\in g} \left\|\Omega_{i,j}^k\right\|_F^2}$ where $\left\|\Omega_{i,j}^k\right\|_F^2$ is the square of the entry-wise $L_2$ norm of the precision sub-matrix $\Omega_{i,j}^k$. By using this penalty item, we ensure that the column pairs in the same group have similar interaction strength. That is, if one column pair in a particular group has a relatively strong interaction (i.e., $\left\|\Omega_{i,j}^k\right\|_F^2$ is large), the other column pairs in this group shall also have a larger interaction strength. In the opposite, if one column pair in a particular group has a relatively weak interaction (i.e., $\left\|\Omega_{i,j}^k\right\|_F^2$ is small), the other column pairs in this group shall also have a smaller interaction strength.

The parameter $\lambda_g$ is used to enforce residue co-evolution consistency in the same group. It is proportional to the conservation level in group $g$. We measure the conservation level using both the square root of the number of aligned families in a group and also the alignment probabilities. In particular, $\lambda_g$ is defined as follows.

$$\lambda_g = \alpha \sqrt{N-1} \sqrt[N-1]{\prod_{n=1}^{N-1} P_n}$$

Where $\alpha$ is a constant (=0.001), N is the number of column pairs in group $g$ and $P_n$ can be interpreted as the average alignment probability between the target family and the auxiliary family n at the two aligned columns belonging to group g. Meanwhile, $P_n$ is calculated as $P_n = P_i P_j$ where $P_i$ and $P_j$ are the marginal alignment probabilities at the two aligned columns. That is, when the two aligned column pairs are conserved, both $P_i$ and $P_j$ are large, so is $P_n$.

Consequently, $\lambda_g$ is large and thus the interaction strength consistency among the column pairs in group g is strongly enforced. In the opposite, if the marginal alignment probability is relatively small, $\lambda_g$ is small. In this case, we shall not strongly enforce the interaction strength consistency among column pairs in this group. By using the conservation level (or alignment quality) to control the correlation of interaction strength, our method is robust to bad alignments and thus, can also deal with protein families similar at different levels.

Note that our formulation (31) differs from the PSICOV formulation only in the last term, which is used to enforce co-evolution pattern consistency among multiple families. Without this term, our formulation is exactly the same as PSICOV when the same $\lambda_1$ is used. We use an ADMM algorithm to solve formulation (31) as the following.

**Estimating precision matrices by Alternating Directions Method of Multipliers (ADMM)**

Computationally, Eqs. (31) and (32) can be solved using almost the same ADMM procedure, so here we explain how to solve Eq. (31) using ADMM (see https://web.stanford.edu/~boyd/papers/pdf/admm_slides.pdf). To estimate the precision matrix for the target protein family, we shall solve the following optimization problem.

$$(P1) \quad max_\Omega f(\Omega) - \lambda_1 \sum_{k=1}^{K} \|\Omega^k\|_1 - \lambda_g \sum_{g=1}^{G} \|\Omega_g\|_2 \quad (33)$$

$$f(\Omega) = \sum_{k=1}^{K} \left( \log|\Omega^k| - \mathrm{tr}(\Omega^k \hat{\Sigma}^k) \right)$$

Where $\Omega$ denote the set $\{\Omega^1, \Omega^2, \dots, \Omega^k\}$. To apply ADMM, we rewrite P1 as a constrained optimization problem by making a copy of $\Omega$ to $Z$, but without changing the optimal solution.

$$(P2) \quad max_{\Omega,Z} f(\Omega) - P(Z) \quad (34)$$

$$P(Z) = \lambda_1 \sum_{k=1}^{K} \|Z^k\|_1 - \lambda_g \sum_{g=1}^{G} \|Z_g\|_2$$

$$s.t \quad \forall k, \quad \Omega^k = Z^k$$

Where $Z$ denote the set $\{Z^1, Z^2, \dots, Z^k\}$. Eq. (34) can be augmented by adding one term to penalize the difference between $\Omega^k$ and $Z^k$ as follows.

$$(P3) \quad max_{\Omega,Z} f(\Omega) - P(Z) - \sum_{k=1}^{K} \frac{\rho}{2} \|\Omega^k - Z^k\|_F^2 \quad (35)$$

$$s.t \quad \forall k, \quad \Omega^k = Z^k$$

P3 is equivalent to P2 and P1, but converges faster due to the added penalty term. Here $\rho$ is a hyper-parameter controlling the convergence rate. Some heuristics methods (Rush, et al., 2010; Sontag, et al., 2011) were proposed for choosing $\rho$. In our implementation, we set $\rho$ to 0.1. On most test cases, our algorithm can converge within 40 iterations. Using a Lagrange multiplier $U^k$ for each constraint $\Omega^k = Z^k$, we obtain the following Lagrangian dual problem.

(P4) $\quad min_U max_{\Omega,Z} f(\Omega) - P(Z) - \sum_{k=1}^{K} (\rho U^k)^T (\Omega^k - Z^k) + \frac{\rho}{2} \|\Omega^k - Z^k\|_F^2 \quad$ (36)

It is easy to prove that P4 is an upper bound of P3. Instead of directly solving P3, we solve P4 iteratively. At each iteration, we fix $U$ and solve the following sub-problem.

(P5) $\quad max_{\Omega,Z} f(\Omega) - P(Z) - \frac{\rho}{2} \|\Omega^k - Z^k + U^k\|_F^2 \quad$ (37)

The sub-gradient of $U$ is $-\rho(\Omega - Z)$, so we may update $U$ by $U + \rho(\Omega - Z)$ and repeat solving P5 until convergence, i.e., the difference between $\Omega$ and $Z$ is small.

To solve P5, we decompose it into the below two sub-problems and then solve them alternatively.

(SP1) $\quad \forall k, \ (\Omega^k)^* = argmax_\Omega \{ f(\Omega) - \frac{\rho}{2} \|\Omega^k - Z^k + U^k\|_F^2 \} \quad$ (38)

(SP2) $\quad Z^* = argmin_\Omega \{ \frac{\rho}{2} \|\Omega^k - Z^k + U^k\|_F^2 \} \quad$ (39)

Meanwhile, SP1 optimizes the objective function (37) with respect to $\Omega$ while fixing $Z$ and $U$. Since in SP1 no two $\Omega^K$ are coupled together, we can split it into $K$ independent optimization sub-problems. SP2 optimizes the objective function with respect to $Z$ while fixing $\Omega$. Next we will show how to solve these two sub-problems efficiently.

Solving SP1. SP1 is a concave and smooth function, so we can solve it by setting its derivate to zero as follows.

$$((\Omega^k)^{-1} - \hat{\Sigma}^k) - \rho(\Omega^k - Z^k + U^k) = 0 \quad (40)$$

Let $M^k = \hat{\Sigma}^k - \rho Z^k + \rho U^k$. Then we have $M^k = {\Omega^k}^{-1} - \rho \Omega^k$. That is, $M^k$ has the same eigenvalues and eigenvectors as $(\Omega^k)^{-1} - \rho \Omega^k$. Since $\Omega^k$ and $(\Omega^k)^{-1}$ share the same eigenvectors, $M^k$ should have the same eigenvectors as $(\Omega^k)^{-1} - \rho \Omega^k$. Let $\delta_i$ and $m_i$ be the ith eigenvalues of matrix $\Omega^k$ and $M^k$, respectively, and let $x_i$ ($\neq 0$) be the corresponding eigenvector. Then we have $M^k x_i = (\hat{\Sigma}^k - \rho Z^k + \rho U^k) x_i$. That is, $m_i x_i = (\delta_i^{-1} - \rho \delta_i) x_i$. So we have $m_i = \delta_i^{-1} - \rho \delta_i$, from which we can solve $\delta_i$ as follows.

$$\delta_i = \frac{-m_i + \sqrt{m_i^2 + 4\rho}}{2\rho} \tag{41}$$

Therefore, we can first do SVD (singular value decomposition) on Mk, and then reconstruct $\Omega^k$ from $\delta_i$ and the eigenvectors of $M^k$. SVD is time-consuming. Since $M^k$ is symmetric and sparse, we can permute its rows and columns to obtain a diagonal block matrix, which can be done within running time linear in the number of non-zero elements in $M^k$. Then we divide $M^k$ it into small sub-matrices and calculate their eigenvalues and eigenvectors separately.

Solving SP2. SP2 is a non-differentiable convex function and we can solve it by setting its sub-gradients to zero. That is, for each k, we have the following equation.

$$\rho(Z_{i,j}^k - A_{i,j}^k) + \lambda_1 \frac{Z_{i,j}^k}{\beta_g} + \lambda_g t_{i,j}^k = 0 \tag{42}$$

Where $A^k = \Omega^k + U^k$, $t_{i,j}^k$ is the derivative of $|Z_{i,j}^k|$ and $\beta_g = \|Z_{i,j}^k\|_2$. Meanwhile, $t_{i,j}^k$ is equal to any value between -1 and 1 when $Z_{i,j}^k$ is 0 and otherwise, to $sign(Z_{i,j}^k)$. To solve a particular $Z_{i,j}^k$ based on Eq. (42), we need to know the value of $\beta_g$, which depends on all the $Z_{i,j}^k$ in group $g$. That is, we cannot solve these K optimization problems separately.

Let $S(x, c) = \max(x - c \times sign(x), 0)$. Eq. (42) can be written as follows.

$$\left(1 + \frac{\lambda_g}{\rho \beta_g}\right) Z_{i,j}^k = S(A_{i,j}^k, \frac{\lambda_1}{\rho}) \tag{43}$$

Taking the square of both sides in Eq. (43) and summing up over all $(i,j,k) \in g$, we have the following equation.

$$\sum_{(i,j,k) \in g} \left(1 + \frac{\lambda_g}{\rho \beta_g}\right)^2 (Z_{i,j}^k)^2 = \sum_{(i,j,k) \in g} S(A_{i,j}^k, \frac{\lambda_1}{\rho})^2 \tag{44}$$

By definition, $\sum_{(i,j,k) \in g}(Z_{i,j}^k)^2 = \beta_g^2$, since $\left(1 + \frac{\lambda_g}{\rho \beta_g}\right)^2$ is independent of $(i,j,k)$, the left hand side of Eq. (44) is equal to $(\beta_g + \frac{\lambda_g}{\rho})^2$. Therefore, we can represent $\beta_g$ as follows.

$$\beta_g = \sqrt{\sum_{(i,j,k) \in g} S(A_{i,j}^k, \frac{\lambda_1}{\rho})^2} - \frac{\lambda_g}{\rho} \tag{45}$$

Plugging Eq. (45) back into Eq. (43), we obtain the value of $Z_{i,j}^k$ as follows.

$$Z_{i,j}^k = S(A_{i,j}^k, \frac{\lambda_1}{\rho})(1 - \frac{\lambda_g}{\rho \sqrt{\sum_{(i,j,k) \in g} S(A_{i,j}^k, \frac{\lambda_1}{\rho})^2}}) \tag{46}$$

**Including predicted probability by supervised learning as prior information**

Compared to single-family EC analysis, our joint EC analysis uses residue co-evolution information from auxiliary families to improve contact prediction. In addition to co-evolution information, sequence profile and some non-evolutionary information are also useful for contact prediction. To make use of them, we first use a supervised machine learning method Random Forests to predict the probability of two residues forming a contact and then integrate this predicted probability as prior into our GGL framework. In particular, our Random Forests model predicts the probability of two residues forming a contact using the following information.

PSI-BLAST sequence profile. To predict the contact probability of two residues, we use their position-specific mutation scores and those of the sequentially adjacent residues.

Mutual information (MI) and its power series. When residue $A$ has strong interaction with $B$ and $B$ has strong interaction with residue $C$, it is likely that residue $A$ also has interaction with $C$. We use the $MI^k$ power series to account for this kind of chaining effect. In particular, we use $MI^k$ where $k$ ranges from 2 to 11 where MI is the mutual information matrix. When there are many sequence homologs, the MI power series are very helpful to medium- and long-range contact prediction. EPAD: a context-specific distance-dependent statistical potential (Zhao and Xu, 2012), derived from protein evolutionary information. The $C_\alpha$ and $C_\beta$ atomic interaction potential at all the distance bins is used. The atomic distance is discretized into bins by 1Å and all the distance>15Å is grouped into a single bin.

**Amino acid physic-chemical properties.**

Some features are calculated on the residues in a local window of size 5 centered at the residues under consideration. In total there are ~300 features for each residue pair.

We trained and selected the model parameters of our Random Forests model by 5-fold cross validation. In total we used about 850 training proteins, all of which have <25% sequence identity with our test proteins. See paper (Wang and Xu, 2013) for the description of the training proteins.

Finally, our GGL formulation with predicted contact probability as prior is as follows.

$$max \sum_{k=1}^{K} \left( log|\Omega^k| - tr(\Omega^k \hat{\Sigma}^k) \right) - \lambda_1 \sum_{k=1}^{K} \|\Omega^k\|_1 - \sum_{g=1}^{G} \lambda_g \|\Omega_g\|_2 - \lambda_2 \sum_{k=1}^{K} \sum_{ij} \frac{\|\Omega_{ij}^k\|_1}{max(P_{ij}^k, 0.3)} \quad (47)$$

Where $P_{ij}^k$ is the predicted contact probability by Random Forests and $max(P_{ij}^k, 0.3)$ is used to reduce the impact of very small predicted probability. Meanwhile, $exp(-\lambda_2 \sum_{k=1}^{K} \sum_{ij} \frac{\|\Omega_{ij}^k\|_1}{max(P_{ij}^k, 0.3)})$ can be interpreted as the prior probability of $\Omega$, which is used to promote the similarity between the precision matrix and the predicted contact probability. Formulation (4) differs from formulation (3) only in the last term. From computational perspective, $\lambda_2 \sum_{k=1}^{K} \sum_{ij} \frac{\|\Omega_{ij}^k\|_1}{max(P_{ij}^k, 0.3)}$ is similar to $\lambda_1 \sum_{k=1}^{K} \|\Omega^k\|_1$, so we can use almost the same computational method to optimize both formulations.

**Alignment of multiple protein families**

To build the alignment of multiple protein families, we employ a probabilistic consistency method in (Do, et al., 2006). To employ this consistency method, we need to calculate the probabilistic alignment matrix between any two protein families. Each matrix entry is the marginal alignment probability (MAP) of two columns, each in one family. In addition to this probability method, we also employed MCoffee (Wallace, et al., 2006) to generate alignment of multiple families.

**Majority voting method for contact prediction**

Majority voting is a simple way of utilizing auxiliary protein families for contact prediction. We first build an alignment of multiple protein families using the methods mentioned above. Then we use PSICOV to predict contact map for each of the related protein families. To determine if there is a contact between any two columns $i$ and $j$ in the target protein family, we use a majority voting based upon the predicted contacts for all the column pairs aligned to the pair $(i,j)$. In addition, we also assign a weight to each family proportional to the number of non-redundant sequence homologs in it. The more NR sequence homologs, the more weight this family carries since usually such a family has higher contact prediction accuracy. In this experiment, each protein family is modeled using a different probability distribution since PSICOV is applied to each of the related families separately.

**Pre-processing and Post-processing**

We employ the same pre- and post-processing procedures as PSICOV to ensure our comparison with PSICOV is fair. Briefly, to reduce the impact of redundant sequences, we apply the same sequence weighting method as PSICOV. In particular, duplicate sequences are removed and columns containing more than 90% of gaps are also deleted. The sequence is weighted using a threshold of 62% sequence identity. We add a small constant (=0.1) to the diagonal of the empirical covariance matrix to ensure it is not singular. Similar to PSICOV and plmDCA, average-product correction (APC) (Jones, et al., 2012) is applied to post-process predicted contacts.

## 6.3 Results

We use two datasets to evaluate the performance of our method. One is a subset of the benchmark used in the PSICOV paper, consisting of 98 Pfam families, each of which has at least one auxiliary family. As shown in Eq. (47), when no auxiliary families are available, our method becomes normal graphical lasso with supervised prediction as prior. By considering only the Pfam families with auxiliary families, we can evaluate the impact of auxiliary families. The other dataset consists of the 123 CASP10 targets, some of which do not have auxiliary families.

**PSICOV dataset**

It is selected from the 150 Pfam (Bateman, et al., 2004; Finn, et al., 2014) families used by PSICOV as benchmark, all of which have solved structures in PDB. To make a fair comparison, we use the same solved structures as PSICOV to calculate native contacts. Only $C_\alpha$ contact prediction results are presented. Similar performance trend is observed for $C_\beta$ contacts. We denote these Pfam families, for which we would like to predict contacts, as the target families. For each target family, we find its related families in Pfam, also called auxiliary families, using HHpred with E-value=10-6 as cutoff. As a result, 98 families have at least one auxiliary family and are used as our test data. We can also relax the E-value cutoff to obtain more distantly-related auxiliary families, but this does not lead to significant accuracy improvement. Among the 98 target families, the average TM-score between the representative solved structures of a target family and of its auxiliary families is ~0.7. That is, the target and auxiliary families are not very close, although they may have similar folds. Even using E-value≤10-17 as cutoff, some target and auxiliary families are only similar at the SCOP fold level.

To ensure that the Pfam database does not miss important sequence homologs, we generate an MSA for each target family by PSI-BLAST (5 iterations and E-value=0.001) and then apply PSICOV to this MSA. Such a method is denoted as PSICOV_b. Since HHblits sometimes may detect sequence homologs of higher quality than PSI-BLAST, we also run HHblits to build an MSA for a target sequence and then examine if this MSA can lead to better prediction or not.

**Methods to be compared**

We compare our method with a few popular EC methods such as PSICOV, Evfold, plmDCA and GREMLIN and a few supervised learning methods such that NNcon and CMAPpro. We use their default parameter settings. Since both plmDCA and GREMLIN use the pseudo-likelihood methods, we run Evfold with the mean field solution instead of the pseudo-likelihood solution to diversify the set of methods to be compared.

There are two alternative strategies to use information in auxiliary families. One is that we can merge a target and its auxiliary families into a single MSA and then apply single-family EC analysis. To test this strategy, we align and merge a target and its auxiliary families into a single MSA using a probabilistic consistency method and MCoffee, respectively, and denote them as Merge_p and Merge_m. The other strategy, denoted as Voting, is that we apply the single-family EC method PSICOV to each of the target and auxiliary families and then apply a majority voting method to predict the contacts in the target family.

We evaluate the top L/10, L/5 and L/2 predicted contacts where L is the sequence length of a protein (family) under prediction. The contact prediction accuracy is defined as the percentage of native contacts in the top predicted contacts. When more predicted contacts are evaluated, the difference among methods decreases since it is more likely to pick a native contact by chance. Contacts are short-, medium- and long-range when the sequence distance between the two residues in a contact falls into three intervals [6,12), [12, 24), and $\geq 24$, respectively. Generally speaking, medium- and long-range contacts are more important, but more challenging to predict.

**Overall performance on the PSICOV test set**

As shown in Table 22, tested on all the 98 Pfam families with auxiliary families, our method CoinDCA outperforms the others when the top L/10, L/5 and L/2 predicted contacts are evaluated, no matter whether the contacts are short-, medium- and long-range. When neither auxiliary families nor supervised learning is used, CoinDCA is exactly the same as PSICOV. Therefore, the results in Table 1 indicate that combining joint EC analysis and supervised learning indeed can improve contact prediction accuracy over single-family EC analysis. We also observe the following performance trends.

1) In terms of contact prediction, the MSAs generated by PSIBLAST or HHblits are not better than those in Pfam.

2) A simple majority voting scheme performs worse than the single-family EC methods. This may be due to a couple of reasons. When a single family is considered, PSICOV may wrongly predict contacts in each family in very different ways, so consensus of single-family results can only identify those highly-conserved contacts, but not those specific to one or few families. In addition, majority voting may suffer from alignment errors.

3) It does not work well by merging the target and auxiliary families together into a single MSA and then applying single-family EC analysis. There are two possible reasons. One is that the resultant MSA may contain alignment errors, especially when the auxiliary families are not very close to the target family. The other is that we cannot use a single Gaussian distribution to model the related but different families due to sequence divergence (at some positions). Since Merge_p performs better than Merge_m, we will consider only Merge_p in the following sections.

PSICOV models the MSA of a protein family using a multivariate Gaussian distribution. This Gaussian assumption holds only when the family contains a large number of sequence homologs. plmDCA and GREMLIN do not use the Gaussian assumption and are reported to outperform PSICOV on some Pfam families. Our method CoinDCA still uses the Gaussian assumption. This test result indicates that when EC information in multiple related families is used, even with Gaussian assumption, we can still outperform the single-family EC methods without using Gaussian assumption.

**Table 22.** Contact prediction accuracy on all the 98 test Pfam families. plmDCA and GREMLIN use the MSAs in the Pfam database while plmDCA_h and GREMLIN_h use the MSAs generated by HHblits.

|  | Short-range | | | Medium-range | | | Long-range | | |
|---|---|---|---|---|---|---|---|---|---|
|  | L/10 | L/5 | L/2 | L/10 | L/5 | L/2 | L/10 | L/5 | L/2 |
| **CoinDCA** | **0.528** | **0.446** | **0.316** | **0.496** | **0.435** | **0.312** | **0.561** | **0.502** | **0.391** |
| **PSICOV** | 0.369 | 0.299 | 0.205 | 0.375 | 0.312 | 0.213 | 0.446 | 0.400 | 0.311 |
| **PISCOV_h** | 0.382 | 0.306 | 0.204 | 0.418 | 0.334 | 0.218 | 0.466 | 0.421 | 0.310 |
| **PSICOV_b** | 0.356 | 0.286 | 0.199 | 0.388 | 0.306 | 0.199 | 0.462 | 0.400 | 0.294 |
| **Merge_p** | 0.316 | 0.265 | 0.183 | 0.303 | 0.246 | 0.178 | 0.370 | 0.328 | 0.253 |
| **Merge_m** | 0.298 | 0.237 | 0.172 | 0.276 | 0.223 | 0.169 | 0.355 | 0.309 | 0.232 |
| **Voting** | 0.343 | 0.232 | 0.184 | 0.405 | 0.280 | 0.168 | 0.337 | 0.353 | 0.275 |
| **plmDCA** | 0.422 | 0.327 | 0.203 | 0.433 | 0.354 | 0.233 | 0.484 | 0.443 | 0.343 |
| plmDCA_h | 0.387 | 0.300 | 0.186 | 0.433 | 0.339 | 0.211 | 0.480 | 0.413 | 0.292 |
| plmDCA_b | 0.381 | 0.301 | 0.184 | 0.431 | 0.338 | 0.210 | 0.478 | 0.421 | 0.289 |
| GREMLIN | 0.410 | 0.312 | 0.220 | 0.401 | 0.332 | 0.225 | 0.447 | 0.423 | 0.329 |
| GREMLIN_h | 0.387 | 0.291 | 0.188 | 0.391 | 0.316 | 0.204 | 0.428 | 0.400 | 0.301 |
| GREMLIN_b | 0.379 | 0.289 | 0.187 | 0.390 | 0.314 | 0.203 | 0.426 | 0.398 | 0.303 |
| Evfold | 0.340 | 0.274 | 0.191 | 0.364 | 0.298 | 0.209 | 0.400 | 0.361 | 0.281 |
| Evfold_h | 0.326 | 0.250 | 0.171 | 0.345 | 0.279 | 0.189 | 0.381 | 0.333 | 0.262 |
| Evfold_b | 0.324 | 0.252 | 0.169 | 0.344 | 0.275 | 0.190 | 0.382 | 0.332 | 0.261 |

**Dependency on the number of sequence homologs**

Our method outperforms the others regardless of the size of a protein family. Similar to (Jones, et al., 2012), we calculate the number of non-redundant sequence homologs in a family (or multiple sequence alignment) by $M_{eff} = \sum_i 1/\sum_j s_{i,j}$ where $i$ and $j$ are sequence indexes and $s_{i,j}$ is a binary variable indicating if two sequences are similar or not. It is equal to 1 if the normalized hamming distance between two sequences is less than 0.3; otherwise, 0. The reason why we use $M_{eff}$ instead of the number of sequences to quantify the information content in an MSA is that there may exist many highly similar homologs in the MSA. Highly similar homologs do not provide more information for co-evolution detection than a single one, so we can only count the number of non-redundant sequence homologs. We divide the 98 test families into 5 groups by $\ln M_{eff}$: [4,5), [5,6), [6,7), [7,8), [8,10), and calculate the average L/10 prediction accuracy in each group. Figure 23 shows that our method performs significantly better than the others regardless of $\ln M_{eff}$. In particular, the advantage of our method over the others is even larger when $\ln M_{eff}$ is small.

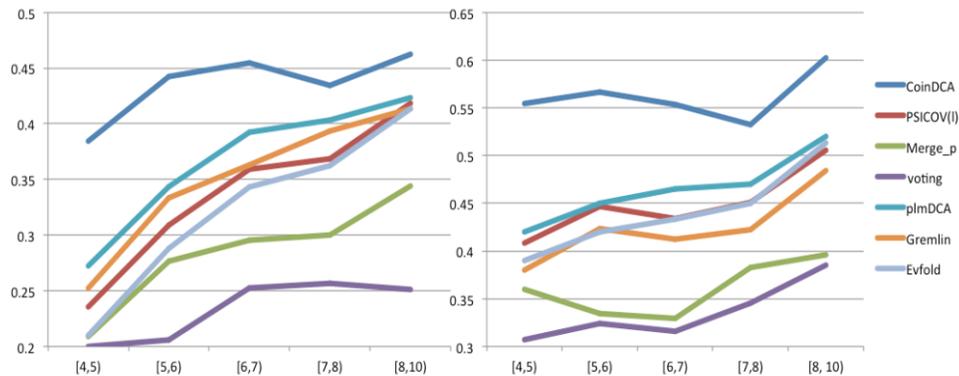

**Figure 23.** (A) Medium-range and (B) Long-range L/10 prediction accuracy with respect to $\ln M_{eff}$.

**Performance and contact conservation level**

For a native contact in the target family, we measure its conservation level by the number of auxiliary families with a contact alignable to this target contact. The 98 test families have conservation levels ranging from 0 to 8, corresponding to non-conserved and highly-conserved, respectively. In particular, a native contact with a conservation level of 0 is target family-specific since it has no support from any auxiliary families. Correct prediction of family-specific contacts is important since they may be very useful to the refinement of a template-based protein model.

Figure 24 (A) and (B) shows the ratio of medium- and long-range native contacts ranked among top L/10 predictions with respect to contact conservation level. Our method CoinDCA ranks many more native long-range contacts among top L/10 than the single-family EC methods PSICOV, plmDCA and GREMLIN regardless of conservation level. CoinDCA has similar performance as the family merging method Merge_p for long-range contacts with conservation level ≥5, but significantly outperforms Merge_p for family-specific contacts. This may be because when the target and auxiliary families are merged together, the signal for highly-conserved contacts is reinforced but that for family-specific contacts is diluted. By contrast, our joint EC analysis method can reinforce the signal for highly-conserved contacts without losing family-specific information.

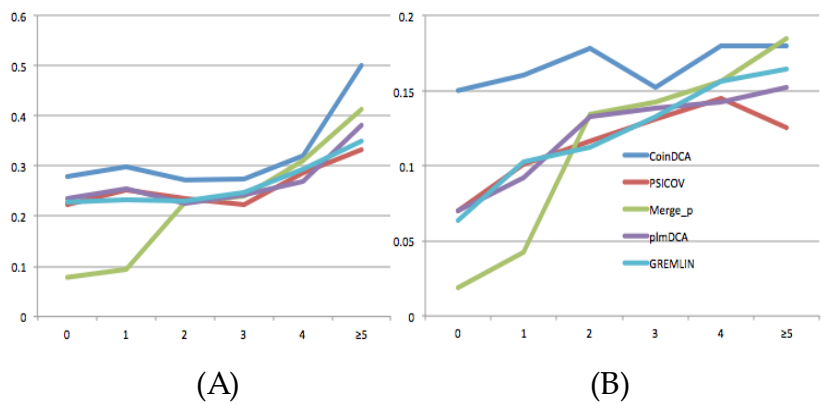

(A)             (B)

**Figure 24.** The ratio (*Y*-axis) of native contacts ranked by a prediction method among top *L*/10 with respect to contact conservation level (*X*-axis) for (A) medium-range and (B) long-range.

**Performance on the CASP10 targets**

In this test we run NNcon, PSICOV, plmDCA, GREMLIN and EVfold locally with default parameters, and CMAPpro through its web server. Again, we run HHpred to search the Pfam database for auxiliary families for each test target.

Meanwhile, 75 of 123 targets have at least one auxiliary family. For those targets without any auxiliary families, our method actually becomes the combination of single-family EC analysis and supervised learning. As shown in Table 23, on the whole CASP10 set, our method CoinDCA again outperforms the others in terms of the accuracy of the top *L*/10, *L*/5 and *L*/2 predicted contacts.

**Table 23.** Contact prediction accuracy on all the 123 CASP10 targets. See supplementary for statistical significance (i.e., P-value).

|  | Short-range | | | Medium-range | | | Long-range | | |
| --- | --- | --- | --- | --- | --- | --- | --- | --- | --- |
|  | L/10 | L/5 | L/2 | L/10 | L/5 | L/2 | L/10 | L/5 | L/2 |
| CoinDCA | **0.517** | **0.435** | **0.311** | **0.500** | **0.440** | **0.340** | **0.412** | **0.351** | **0.279** |
| PSICOV | 0.234 | 0.191 | 0.140 | 0.310 | 0.259 | 0.192 | 0.276 | 0.225 | 0.168 |
| plmDCA | 0.264 | 0.218 | 0.152 | 0.344 | 0.289 | 0.214 | 0.326 | 0.280 | 0.213 |
| NNcon | 0.499 | 0.399 | 0.275 | 0.393 | 0.334 | 0.226 | 0.239 | 0.188 | 0.001 |
| GREMLIN | 0.256 | 0.212 | 0.161 | 0.343 | 0.280 | 0.229 | 0.320 | 0.278 | 0.159 |
| CMAPpro | 0.437 | 0.368 | 0.253 | 0.414 | 0.363 | 0.276 | 0.336 | 0.297 | 0.227 |
| EVfold | 0.193 | 0.165 | 0.130 | 0.294 | 0.249 | 0.188 | 0.257 | 0.225 | 0.171 |

We also divide the 123 CASP10 targets into five groups according to ln M$_{eff}$: (0,2), (2,4), (4,6), (6,8), (8,10), which contain 19, 17, 25, 36 and 26 targets, respectively. Meanwhile, $M_{eff}$ measures the number of non-redundant sequence homologs available for a target protein under prediction. We then calculate the average medium- and long-range contact prediction accuracy in each group. Figure 23 clearly shows that the prediction accuracy increases with respect to $ln\,M_{eff}$ and that our method outperforms the others on all the 5 intervals of $ln\,M_{eff}$. In particular, our method works much better than the single-family EC analysis methods when $ln\,M_{eff}$ <8.

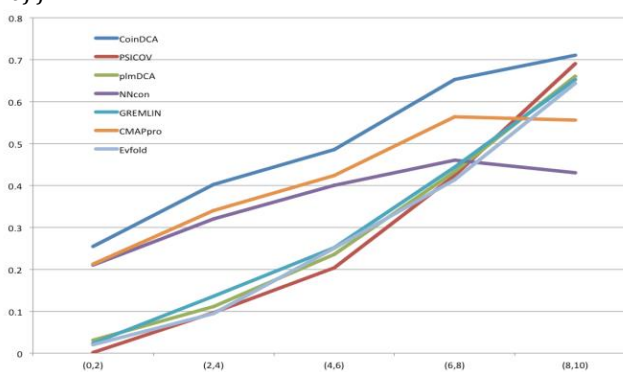

**Figure 23.** The relationship between prediction accuracy and $ln\,M_{eff}$. X-axis is the $ln\,M_{eff}$ value and Y-axis is the mean accuracy of top L/10 predicted contacts in the corresponding CASP10 target group. Only medium- and long-range contacts are considered.

## 6.4 Discussion

In this chapter I has presented a GGL method to predict contacts by integrating joint EC analysis and supervised machine learning. Evolutionary coupling (EC) analysis and supervised learning are currently two major methods for contact prediction, but they use different information sources. Our joint EC analysis predicts contacts in a target family by analyzing residue co-evolution information in a set of related protein families which may share similar contact maps. In order to effectively integrate information across multiple families, we use GGL to estimate the joint probability distribution of multiple related families by a set of correlated Gaussian models. Experiments show that the combination of joint EC analysis with supervised machine learning can significantly improve contact prediction, and that our method even outperforms single-family EC analysis on protein families with a large number of sequence homologs. We have also shown that contact prediction cannot be improved by a simple method, such as family merging and majority voting of single-family EC analysis results. These simple methods may improve prediction for highly-conserved contacts at the cost of family-specific contacts.

Our method can be further improved. For example, similar to GREMLIN and plmDCA, we may relax the Gaussian assumption to improve prediction accuracy. This paper uses an entry-wise $L2$ norm to penalize contact map inconsistency among related protein families. There may be other penalty functions that can more accurately quantify contact map similarity between two families as a function of sequence similarity and thus, further improve contact prediction. It may further improve contact prediction by integrating other supervised learning methods such as CMAPpro, NNcon and DNcon or even other EC methods into our GGL framework.

In this work we use Pfam to define a protein family because it is manually-curated and very accurate. There are also other criteria to define a protein family. For example, SCOP defines a protein family based upon structure information and thus, classifies protein domains into much fewer families than Pfam. In our experiment, the average structure similarity, measured by TM-score, between a target (Pfam) family and its auxiliary (Pfam) families is only around 0.7. That is, many auxiliary families are not highly similar to its target families even by the SCOP definition. Indeed, some auxiliary families are only similar to the target family at the SCOP fold level. That is, even a remotely-related protein family may provide information useful for contact prediction.

We can further extend our method to predict contacts of all the protein families simultaneously, instead of one-by-one, by joint EC analysis across the whole protein family universe. First we can use a graph to model the whole Pfam database, each vertex representing one Pfam family and an edge indicating that two families may be related. Then we can use a graph of correlated GGMs to model the whole Pfam graph, each GGM for one vertex. The GGMs of two vertices in an edge are correlated together through the alignment of their respective protein families. By this way, the residue co-evolution information in one family can be passed onto any family that is connected through a path. As such, we may predict the contacts of one family by making use of information in all the path-connected families. By enforcing this global consistency, we should be able to further improve EC analysis for contact prediction. However, to simultaneously estimate the parameters of all the GGMs, a large amount of computational power will be needed.

# Chapter 7    Conclusion and Future Work

The thesis is aimed to solve the template-based protein structure prediction problem by improving the quality of protein alignment. We have developed a new alignment method that align two families through alignment of two Markov Random Fields (MRFs), which model the multiple sequence alignment (MSA) of a protein family using an undirected general graph in a probabilistic way. The node alignment potential can handle the complex relationship between various kinds of features and alignment alignments (Chapter 4, Chapter 5). The edge alignment potential can integrate both supervised-learning-based and evolutionary-coupling-based (Chapter 6) interaction strength to quantify the global similarity between pairs of residues on two proteins to be aligned. Experiments show that our different alignment methods can generate more accurate alignments and is also much more sensitive than other state-of-the-art methods.

**Building Alignments suitable for 3D modeling of protein structure.**

For template-based protein structure prediction, there are correlations between the alignment accuracy and quality of the model recovered using the alignment. However, it is not always the case that the more accurate of the alignment between target and template proteins, the higher quality the recovered model has. The reason is that the residues on the primary sequence are not equally important for structure prediction. Aligning some of the most crucial residues correctly to the template is more important than aligning others correctly. For example, several shifts of alignment between two beta sheets (helix) from two proteins might not influence the final model quality very much while missing aligning one "hinge" residue might lead to completely wrong domain orientation in the recovered model. From a machine learning perspective, it is very hard to design a computational model that can achieve zero training error on the training and validation sets. This could be caused by: 1) The features we use is not discriminative enough. 2) The computational model is not powerful enough to capture all the correlations among features and samples. From another perspective, we usually add some penalty terms (prior probability) to prohibit zero error to overcome over-fitting. If we know we have to make some mistakes in the training and testing data sets (suppose they are from the same distribution), we do not want our model to miss align the most crucial residues for template-based protein structure prediction. Here we propose a novel alignment method that will produce alignments suitable for model recovery.

To overcome this problem, we use a newly developed machine learning model called HCsearch to find the alignment directly maximizing the model quality. The framework uses a search procedure guided by a learned heuristic $H$ to uncover high quality candidate alignments and then uses a separate learned cost function $C$ to select a final alignment among those candidates.

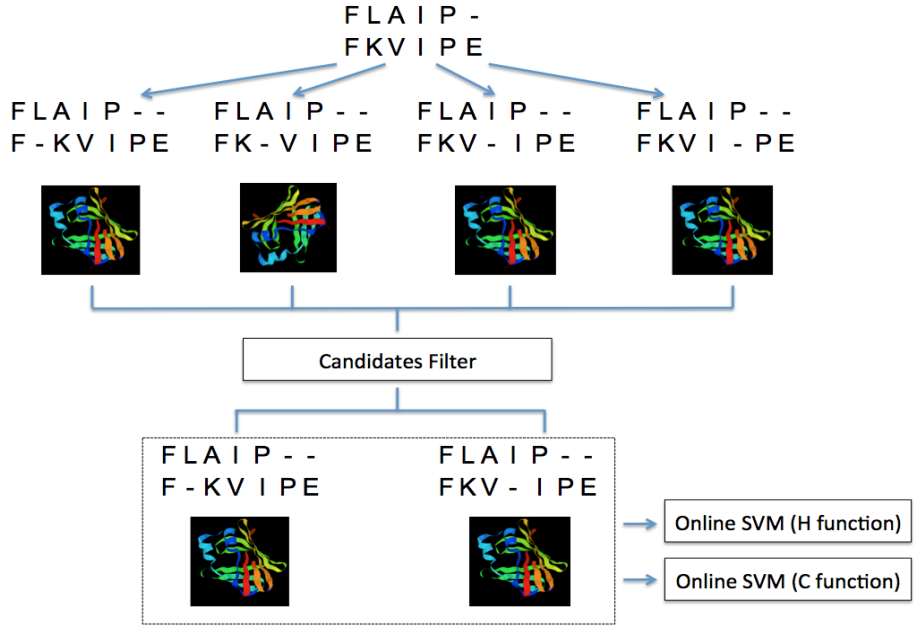

**Figure 24.** HCsearch based alignment method.

Given a target protein $S$ and a template protein $T$ and their alignment $A_0$, start from $A_0$ we can generate $M$ candidate alignments $\{A_1, A_2, \ldots, A_M\}$. We can recovery $M$ models using 3D modeling software corresponding to these $M$ different alignments. During the training process, since we know the native structure of $S$ we can calculate the model quality for each these $M$ models. We then collect all the pairwise ranking decisions of these $M$ model quality. Ties are broken using a fixed arbitrator. The aggregate set of ranking examples collected over all the training examples is then given to a learning algorithm to learn the weights of the heuristic function. In this work we use the margin-scaled variant of the online Passive-Aggressive algorithm (Doppa, et al., 2013). Let $A_{best}$ to be the alignment with the highest model quality from $\{A_1, A_2, \ldots, A_M\}$. We then let $A_0 = A_{best}$ then repeat the above process for $N$ times. We then use the partial relationship between these $N$ best alignments to train another ranking model using the same online learning algorithm. The first ranking function is called Heuristic function ($H$ function) and the second ranking is called the Cost function ($C$ function). The pipeline of the algorithm is shown in Figure 24.